\documentclass[useAMS,usenatbib]{mn2e}

\input{psfig.sty}
\usepackage{graphicx}
\usepackage{epsfig}
\usepackage{amssymb}
\usepackage{amsmath}
\usepackage{amsfonts}
\usepackage{txfonts}

\title[Non-Gaussian Error Bars in Galaxy Surveys (Part 1)]{Non-Gaussian Error Bars in Galaxy Surveys (Part 1)}
\author[Joachim Harnois-D\'{e}raps and Ue-Li Pen]{Joachim Harnois-D\'{e}raps$^{1,2}$ 
\thanks{E-mail: jharno@cita.utoronto.ca} and Ue-Li Pen$^{1}$ \thanks{E-mail: pen@cita.utoronto.ca} \\
$^{1}$Canadian Institute for Theoretical Astrophysics, University of
Toronto, M5S 3H8, Canada\\
$^{2}$Department of Physics, University of Toronto, M5S 1A7, Ontario,  Canada}

\begin{document}

\date{\today}

\pagerange{\pageref{firstpage}--\pageref{lastpage}} \pubyear{2011}

\maketitle

\label{firstpage}

\begin{abstract}
We propose a method to estimate non-Gaussian error bars on the matter power spectrum  from galaxy surveys
in the presence of non-trivial survey selection functions.  The estimators are often
obtained from formalisms like FKP and PKL,  which rely on the assumption that the underlying field is Gaussian.
The Monte Carlo method is more accurate but involves the tedious process of running and cross-correlating a large
number of N-body simulations, in which the survey volume is embedded. 
From 200 N-body simulations, we extract a non-linear covariance matrix as a function
of two scales and of the angle between two Fourier modes.
All the non-Gaussian features of that matrix are then simply parametrized in terms of
a few fitting functions and Eigenvectors.
We furthermore develop a fast and accurate strategy that combines our parameterization with
a general galaxy survey selection function,
and incorporate non-Gaussian Poisson uncertainty.
We describe how to incorporate these two distinct non-Gaussian contributions into a typical analysis pipeline,
and apply our method with the selection function from the 2dFGRS.
We find that the observed Fourier modes correlate at much larger scales  than that predicted by both FKP  formalism or by
pure N-body simulation in a ``top hat''  selection function.
In particular, the observed Fourier modes are already $50$ per cent correlated at $k \sim  0.1 h\mbox{Mpc}^{-1}$,
and the non-Gaussian fractional variance on the power spectrum ($\sigma^{2}_{P}/P^{2}(k)$) is about a factor of $3.0$  
larger than the FKP prescription. At $k \sim  0.4 h\mbox{Mpc}^{-1}$,  the deviations are an order of magnitude.
\end{abstract}


\section{Introduction}
\label{sec:intro}

With new galaxy surveys probing a larger dynamical range of our Universe, our ability to constrain 
cosmological parameters is improving considerably. 
In particular, one of the most important goal of modern cosmology is to  
understand the nature of dark energy  \citep{2006astro.ph..9591A}, 
a challenging task since there are currently no avenues to observe it directly. 
It is however  possible to probe its dynamics via its equation of state $\omega$, which
enters in the Friedmann equation that governs the expansion of the Universe.
Among different ways $\omega$ can be measured, the detection of the baryonic acoustic oscillations (BAO) dilation scale \citep{  Eisenstein:2005su, 2006PhRvD..74l3507T,2006A&A...449..891H,2007MNRAS.381.1053P,2011arXiv1105.2862B}
is one of the favorite,  both because of the low systematic uncertainty and the potentially high statistics on can achieve
with current \citep{1990ApJS...72..433H,2000AJ....120.1579Y,2003astro.ph..6581C,2010MNRAS.401.1429D} and future galaxy surveys  
\citep{2006astro.ph..6104P,2008PhRvD..78d3514A, 2009arXiv0902.4680S, 2009arXiv0912.0201L, 2009ApJ...691..241B,  2010ASPC..430..266B}. 

The strength of the BAO technique relies on an accurate and precise measurement of the matter power spectrum,
whose uncertainty is propagated on to the dark energy parameters via a Fisher matrix   \citep{Tegmark:1997rp}. 
It is thus of the utmost importance to have optimal estimators of both the mean and the  uncertainty of the power spectrum  to start with.
The prescription to construct an estimator for the power spectrum of a Gaussian random field, 
in a given galaxy survey, was pioneered by Feldmann, Kaiser and Peacock \citep{1994ApJ...426...23F} 
(FKP for short). It states that the survey selection function effectively couples Fourier bands that are otherwise 
independent, and that  the underlying power should then be deconvolved \citep{2011PThPh.125..187S}. 
This technique has been used in power spectrum measurement such as \citep{1994ApJ...426...23F,  2001MNRAS.327.1297P,  2005MNRAS.362..505C,
2006A&A...449..891H,2010MNRAS.406..803B}.
Although it is fast, the error bars between the bands are correlated, plus it has the undesired tendency 
to smear out the underlying power spectrum, which can effectively reduce the signal-to-noise ratio
in a BAO measurement. In that sense, the FKP power spectrum is said to be {\it suboptimal}.

The band correlation induced by the FKP prescription can be removed 
by an Eigenvector decomposition of the selection function, following the Pseudo Karhunen-Lo\`{e}ve formalism \citep{1996ApJ...465...34V}(PKL). 
This was used in the analysis of the SDSS data \citep{2006PhRvD..74l3507T} and is the most optimal (i.e. loss-less) estimator of a convolved Gaussian random field,
as understood from the information theory point of view.
It is nevertheless a well known fact that this Gaussian assumption about the field is only valid in the linear regime,
since the non-linear gravitational collapse of the density  effectively couples different Fourier modes together \citep{ 1999MNRAS.308.1179M, 2005MNRAS.360L..82R}, 
and the phases of the modes are no longer random \citep{2000Natur.406..376C}.
Both the FKP and PKL prescriptions, by their Gaussian treatment, do not take into account  
the intrinsic non-linear coupling of the Fourier modes. 
It follows from this that for both methods, the measured power spectrum is suboptimal
and the error bars are systematically biased.
Although the bias is usually small, it causes a problem 
when estimating derived quantities that need to be measured with a percent level accuracy.  

For instance, the observed BAO signal sits right at the transition between the linear and the non-linear regime, 
therefore the optimal estimator of the power spectrum must incorporate the non-linear modes. 
In particular, constraints on dark energy from BAO measurements require an accurate 
measurement of the matter power spectrum covariance matrix.
Under the FKP and PKL formalisms, the covariance matrix is biased as it tends to 
underestimate the uncertainty and the amount of correlation between the power bands. 
Alternative ways of estimating the error, i.e. methods  that involve mock catalogs, 
do model these non-linear dynamics, but it is not clear that the results are precise enough to measure four-points statistics,
and we rather rely on accurate N-body simulations.

Even more relevant is the recent realization  that an optimal, i.e. non-Gaussian, 
estimate of the BAO dilation scale requires a precise measurement of the {\it inverse} of the matrix,
which is challenging due the noisy nature of the forward matrix. 
It was nevertheless shown that, first, a suboptimal measurement of the 
power spectrum should be accompanied with error bars that treat the 
mean as suboptimal by consistency. These error bars differ from the naive Gaussian 
approximation by a significant amount \citep{2011arXiv1106.5548N}.
Second, it was shown in the same paper that an optimal measurement of the mean power spectrum
could lead to an optimal measurement of the error, which differs from suboptimal measurements
by a few percent. To achieve this, however, one needs to include the non-linear dynamics
at a high precision, and  needs in addition a strategy to cope with the noise present in the 
covariance matrix.    

There are thus two aspects of the analysis that need to be extended from the above conclusions.
On one hand, one would like to get rid of the aforementioned bias on the estimated uncertainty in actual data analyses,
in which non-trivial survey selection function play a complex role. 
On the other hand, one would like to measure optimal error bars on the matter power spectrum of (non-Gaussian) galaxy fields.
In this paper, we address the first issue, and provide a strategy to measure suboptimal but unbiased error bars on the power spectrum of galaxy surveys.
It should be mentioned that our method could be put in conjunction
with the PKL formalism to address the second issue, hence optimal measurement
of both the power spectrum and of its error are now within reach. 

When constructing an estimator of the covariance matrix that corresponds to the sensitivity of a particular survey,
the convolution with the survey  selection function is one of the most challenging part.
Whereas the convolution of the underlying power spectrum can be operated with angle averaged quantities, 
the convolution of the covariance matrix must be done in 6 dimensions,
since the underlying covariance is not isotropic: Fourier modes with smaller angular separations are more correlated
than those with larger angles \citep{2002MNRAS.337..488C,2002PhR...367....1B}. 
The first challenge is to measure accurately this angular correlation, 
which is also scale dependent. Neither second order perturbation theory nor log-normal densities have been shown to 
calculate this quantity accurately, hence we must therefore rely on N-body simulations.
This requires a special approach,  since a naive pair counting of all  Fourier modes in the four-point function, at a given angle, would take forever to compute.
The second challenge comes from the 6-dimensional convolution of the covariance matrix with the survey function.
This is a task that current computer clusters cannot solve by brute force, 
so we must find a way to use symmetries of the system
and reduce the dimension of the integral. The fact is that the underlying covariance really depends only on three variables:
two scales and the relative angles between the two Fourier modes. 
Moreover, it turns out, as we describe in section \ref{sec:Cl}, that it is possible to express this matrix into a set of 
multipoles, each of which can further be decomposed into a product of Eigenvectors. 
This effectively factorizes the three dimensions of the covariance, hence the convolution
can be broken down into smaller pieces. By doing so, the non-Gaussian calculation is within reach, 
and we present in this paper the first attempt at measuring deviations from Gaussian calculations,  including both 
Poisson noise and a survey selection function. 
In short, the main ideas of this paper can be condensed as follow:
\begin{enumerate}
\item { The underlying non-linear covariance matrix of the matter power spectrum exhibits many non-Gaussian
         features in the trans- and non-linear regimes. First, the diagonal elements of the
         angle-averaged covariance grow stronger, and correlation across different scales 
         becomes important. Second, Fourier modes with similar (or identical) magnitudes correlate
         more if the angle between them is small. }
\item { It is possible to model all of these non-Gaussian aspects 
       (including the dependence on the angle between the two Fourier modes) with a small number of simple functions.} 
\item{  With such a parameterization, it is possible, for the first time, to solve the six-dimensional
        integral that enters the convolution of the covariance of the power spectrum with
        the galaxy survey selection function.}
\end{enumerate}
Concerning the second point, the parameters that best fit our measurements are provided in section 
\ref{sec:factorization}, but these are separately testable, and could be verified by other groups and with other ways.
These are anyway expected to change when one uses haloes instead of particles.
The third point is, however, a straight forward recipe that is robust under possible changes of best-fitting parameters,
and provides, assuming that the input parameters are correct, an unbiased measurement 
of the non-Gaussian uncertainty of the matter power spectrum.

As indicated by the title, this paper is the first part of a general strategy  that aims at constructing an unbiased,
non-Gaussian estimator of the uncertainty on the matter power spectrum measured in galaxy survey.
The second part, which we thereafter refer to as  HDP2 (in preparation), exploits the fact that the 
measurement of the $C(k,k',\theta)$ matrix provides a novel handle at measuring  $C(k,k')$:
the two quantities are related by a straight forward integration over $\theta$.
As shown in a later section of the current paper, it turns out that the main contribution to $C(k,k')$
comes from small separation angles, while larger angles are noise dominated.
 It is thus possible to perform a noise weighted integral, which results in a more optimal measurement of $C(k,k')$ and of its error bars,
 compared to direct or bootstrap sampling.
It is therefore  possible to extract an accurate non-Gaussian error bar on the power spectrum
with a much smaller number of realizations.
This opens the door for a measurement of a non-Gaussian covariance matrix directly  from the data (i.e. an internal  estimate),
a significant step forward in the error analysis of galaxy surveys.
In that second part, we will first test the improvement of the noise-weigthed technique 
on the same 200 simulations. 
We will then attempt to extract the mean and the error on all the elements of the $C(k,k')$ matrix from 
a mock survey, that will consist of a handful of simulations.

Back to the current paper, our first objective is thus to measure the covariance of the power spectrum between various scales and angles,
and organize this information into a compact matrix, $C(k,k',\theta)$.
We describe how we solve this problem in a fast way, which is based on a series of fast Fourier transforms  and that
can be run in parallel on a large number of computers. 
We found that the angular dependence, at fixed scales ($k \ne k'$), is rather smooth,
it agrees with analytical predictions in the linear regime, but deviates importantly from Gaussianity for smaller scales.
The dependence is somehow similar when the two scales are identical, up to an delta function for vanishing angles.
We also found that, once projected on to a series of Legendre polynomials, it take very few multipoles
to describe the complete original function. We perform this transform for all scale combinations and group the results
in terms of the multipole moments.  
We also provide a recipe to reconstruct the original covariance matrix, with the angular dependence, from a handful of fitting functions.
This is another advantage of the method we propose here:
it can be separately tested and improved.

Our second objective is to provide a general method to combine this  $C(k,k',\theta)$ with a survey selection function and
non-Gaussian Poisson noise, and hence allow the extraction of non-Gaussian error bars on the measured power spectrum. We test our technique on the publicly available 2dFGRS selection
functions \citep{2002MNRAS.336..907N} and find that there is a significant departure between the Gaussian and non-Gaussian treatment. 
In particular, the fractional error of the power spectrum ($\sigma^{2}_{P}/P^{2}(k)$) at $k\sim  0.1 h\mbox{Mpc}^{-1}$
is about a factor of 3.0 higher in the non-Gaussian analysis, and the departure reaches an order of magnitude 
by $k\sim  0.4h\mbox{Mpc}^{-1}$.
The method proposed here can be also applied to other kinds of BAO experiments, 
including intensity mapping from the emission of the 21 cm line by neutral Hydrogen \citep{2006astro.ph..6104P,2008AIPC.1035..303L,2009arXiv0902.4680S},
or Lyman-$\alpha$ forests surveys \citep{2007PhRvD..76f3009M,2011MNRAS.415.2257M}.
We did not, however, include the effect of redshift distortions, 
and focused our efforts on dark matter density fields obtained from simulated particles.
An improved version of this work would include both of these effects, however.


This paper is organized as follow:
In section \ref{sec:FKP}, we briefly review the FKP method, and describe how to estimate
non-Gaussian error bars in realistic surveys, given a previous knowledge of $C(k,k',\theta)$.   
We then lay down the mathematical formalism that describes how we extract this quantity
from simulated density fields  in section \ref{sec:formalism}.
Section \ref{sec:tests} contains the results from sanity checks and null tests 
that were performed to validate our method, and briefly describes our N-body simulations.
We present our measurement of $C(k,k',\theta)$ in section \ref{sec:rho},  
and describe the multipole decomposition in section \ref{sec:Cl}.
In section  \ref{sec:factorization}, we further simplify the results by extracting 
the principal Eigenvectors  
and provide fitting formulas to reconstruct easily the full covariance matrix.
Section \ref{sec:surveys} contains results of applying our method for a set of simple selection functions.
We finally discuss some implications and extensions of our methods in \ref{sec:discuss},
and conclude in section \ref{sec:conclusion}.

\section{Matter Power Spectrum from Galaxy Surveys}
\label{sec:FKP}

In this section, we quickly review the general FKP method, which is commonly
used in data analysis \citep{1994ApJ...426...23F, 2001MNRAS.327.1297P, 2010MNRAS.406..803B}. 
We then point out some of the major the flaws of such techniques when measuring the uncertainty, 
and describe how non-Gaussian error bars could be estimated in principle.
Before moving on, though, we first lay down the conventions used throughout the paper.
The reader familiar with the FKP method may skip to section \ref{subsec:FKP_cov}.

A continuous density field $\delta({\bf x})$ is related to its
Fourier transform $\delta({\bf k})$ by
\begin{eqnarray}
	\delta({\bf k}) = \int\delta({\bf x})e^{i{\bf k}\cdot{\bf x}}d^{3}x
\end{eqnarray}
where ${\bf k}$ is the wave number corresponding to a given Fourier mode.
The power spectrum $P({\bf k})$ of the field is defined as:
\begin{eqnarray}
	\langle \delta({\bf k}) \delta^{*}({\bf k'})\rangle = (2\pi)^{3}P({\bf k})\delta_{D}({\bf k}-{\bf k'})
\end{eqnarray}
and is related to the mass auto-correlation function by :
\begin{eqnarray}
         \xi({\bf x}) = \frac{1}{(2\pi)^{3}}\int e^{-i{\bf k}\cdot{\bf x}} P({\bf k})d^{3}k
         \label{eq:xi}
\end{eqnarray}
In the above expressions, the angle brackets refer to a volume average in Fourier space, and 
$\delta_{D}({\bf k})$ stands for the Dirac delta function.

\subsection{The optimal estimator of the power spectrum}
\label{subsec:FKP_ps}
The power spectrum of the matter field contains a wealth of information
about the cosmic history and the principal constituents of the Universe.
Unfortunately, it is not directly detectable, since our observations are 
subject to cosmic variance, detection noise, light to mass bias, redshift
distortions and incomplete sky surveys.
The FKP method provides an optimal estimator of the matter power spectrum $P(k)$
under the assumption that the density field is Gaussian.
It is formulated in terms of the survey selection function $W({\bf x})$, the galaxy number density $n$,
the dimensions $(n_{x}, n_{y}, n_{z})$ of the grid where the Fourier transforms are performed,
and the actual number count per pixel $n({\bf x})$.
All the following calculations can be found in \citep{1994ApJ...426...23F}, 
and are included  here for the sake of completeness.

The first step is to construct series of weights $w({\bf x})$ as
\begin{eqnarray}
w({\bf x}) = \frac{1}{1 + W({\bf x})N_{c} n P_{0}} = \frac{1}{1+\bar{n} P_{0}}
\label{eq:weights}
\end{eqnarray}
where $N_{c} = n_{x} n_{y} n_{z}$, $\bar{n}$ is the mean galaxy density  and $P_{0}$ is a characteristic amplitude of the  power spectrum at the scale we want to measure.
Since the latter is not known {\it a priori}, it is usually obtained from a theoretical model, and sometimes updated iteratively. 
The selection function is also normalized such that $\sum_{{\bf x}}W({\bf x})  = 1$.

The optimal estimator of the power spectrum, $P_{est}({\bf k})$, is obtained first by re-weighting each pixel by the weights in [Eq. \ref{eq:weights}],
then by subtracting from the result a random catalog with the same selection function, weights and number of objects $N$.
After taking the expectation value of the results, the 2-points statistics of the pixel counts becomes 
\begin{eqnarray}
\langle n({\bf x})n({\bf x'}) \rangle = \tilde{n} \tilde{n'} (1 + \xi({\bf x} - {\bf x'}))  + 
\bar{n}\delta_{D}({\bf x} - {\bf x'})
\label{eq:FKP_2pt}
\end{eqnarray}
where $\bar{n}$ is the mean density in the patch over which the average is performed. 
The Fourier transform is then given by
\begin{eqnarray}
\langle P_{est}({\bf k})  \rangle  = \frac{|n({\bf k}) - N W({\bf k})|^{2}  - N\sum_{{\bf x}}W({\bf x})w^{2}({\bf x}) }{N^{2}N_{c}\sum_{{\bf x}}W^{2}({\bf x})w^{2}({\bf x})}
\label{eq:FKP_ps}
\end{eqnarray}
where denominator is a convenient normalization.
This measured power is aliased by the grid mass assignment scheme, and should be divided by the appropriate function \citep{2005ApJ...620..559J}.

What this estimator measures is not the underlying power spectrum $P(k)$, 
but a convolution with the survey selection function: 
\begin{eqnarray}
\langle P_{est}({\bf k})  \rangle  =\frac {\sum_{{\bf k'}} P({\bf k'}) |W({\bf k} - {\bf k'})|^{2}}{N_{c}\sum_{{\bf x}}W^{2}({\bf x})w^{2}({\bf x})}
\label{eq:P_est}
\end{eqnarray}
It ideally needs to be deconvolved, an operation that is not always possible.

For many survey geometries, the convolution  effectively transfer power
across different bins which are uncoupled to start with \citep{2006PhRvD..74l3507T}.
As mentioned previously, the PKL prescription also assumes that the density field is Gaussian,
but rotates into a basis in which the bins are decoupled.
In that sense, the PKL technique is more optimal than the FKP, unless the selection
function is close to a ``top hat'', in which case the induced mode coupling vanishes. 
Both case, however, rely on the fundamental
assumption that the underlying density field is Gaussian, 
which is known to be inaccurate in the trans- and non-linear regime,
where one still wants an accurate measure of the power spectrum 
for a BAO analysis. 
Obtaining accurate error bars is a requirement for optimal analyses,
and we shall examine how these are usually obtained. 

\subsection{The FKP covariance matrix}
\label{subsec:FKP_cov}

The covariance matrix of the angle averaged power spectrum is a four point function that contains 
the information about the band error bars, and possible correlation between them.
As mentioned earlier, it is required for many cosmological parameter studies. 
It is generally obtained from the power spectrum as 
\begin{eqnarray}
C(k,k')  = \langle \Delta P(k) \Delta P(k')\rangle
\label{eq:cov_th}
\end{eqnarray}
where $\Delta P(k)$ refers to the fluctuations of the measured values about the mean, 
which is ideally obtained from averaging over many realizations. 
In a typical galaxy survey, such independent realizations are obtained
by sampling well separated patches of the sky. 
Because of the cost of such an operation, the number of patches is usually very small.
The covariance matrix is thus not resolved  from the data,
and the error bars are obtained with external techniques, i.e. from mock catalogs\footnote{We post-pone 
the discussion of mock catalogs until the next section}, or directly from Gaussian statistics (see HDP2  for a prescription
that overcomes this challenge).
For a uniform (top-hat) selection function, the Gaussian covariance matrix is estimated as:
\begin{eqnarray}
C^{Gauss}(k, k')  =  \frac{2}{N(k)}(P(k) + P_{shot})^{2}\delta_{kk'}
\label{eq:Gauss_Cov}
\end{eqnarray}
where $P_{shot}  = 1/n$ and $N(k)$ is the number of Fourier modes that enters in the measurement of $P(k)$.
In the ideal scenario of perfect spherical symmetry and resolution,  
$N(k)  =4\pi k^{2} \Delta k \left(\frac{L}{2 \pi}\right)^{3}$, 
with $\Delta k$ being the width of the $k$-band. 
The Kronecker delta function ensures that
there is no correlation between different modes, an inherent  property of Gaussian random fields.
This equation can be easily be modified to deal with measurements without angle averaging.

The FKP prescription provides a generalization of [Eq. \ref{eq:Gauss_Cov}] for the case 
where the selection function varies across the volume. 
It is obtained from [Eq. \ref{eq:FKP_ps}] and given by
\begin{eqnarray}
C^{FKP}({\bf k}, {\bf k'})  =  \frac{2}{N({\bf k})N({\bf k'})}\sum_{{\bf k},{\bf k'}}|P Q({\bf k} - {\bf k'}) + S({\bf k} - {\bf k'})|^{2}
\label{eq:FKP_Cov}
\end{eqnarray}
where 
\begin{eqnarray}
Q({\bf k})=\frac{\sum_{{\bf x}} W^{2}({\bf x})w^2({\bf x}) \mbox{exp}(i{\bf k}{\bf x})}{\sum_{{\bf x}} W^{2}({\bf x})w^2({\bf x})}
\label{eq:FKP_Q}
\end{eqnarray}
\begin{eqnarray}
S({\bf k})= \left(\frac{1}{n N_{c}}\right)\frac{\sum_{{\bf x}} W({\bf x})w^2({\bf x}) \mbox{exp}(i{\bf k}{\bf x})}{\sum_{{\bf x}} W^{2}({\bf x})w^2({\bf x})}
\label{eq:FKP_S}
\end{eqnarray}
In [Eq. \ref{eq:FKP_Cov}], $P$ is taken to be the mean of the power spectrum at separation ${\bf k} - {\bf k'}$.
Because the selection functions are usually quite compact about ${\bf k}  = 0$, that approximation
is reasonable for Gaussian fields. 
Also, [Eq.\ref{eq:Gauss_Cov}] can be recovered by setting $W({\bf x})  =  1/N_{c}$. 

\subsection{Non-Gaussian error bars}
\label{subsec:non_gauss_error}


As mentioned in the last section, it is necessary to have access to many realization of the matter field in order to measure 
a non-Gaussian covariance matrix of power spectrum. This could in principle be done from data across many different patches in the sky,
but even then, we have only one sky to resolve the largest modes, which would therefore be dominated by cosmic variance.
Not to mention the cost and time involved in measuring many large but disconnected volumes.
Fortunately, N-body simulations are now accurate and fast enough to generate large numbers of  measurements of the matter power spectrum.
Since they model the non-linear dynamics of structure growth, the density fields they generate are non-Gaussian.
The covariance matrix constructed from a high number of simulations indeed shows a correlation across different scales in the non-linear regime
\citep{1999MNRAS.308.1179M,2005MNRAS.360L..82R, 2009ApJ...700..479T,2011arXiv1106.5548N}. 

Although much more representative of the underlying covariance, such matrices are hard to incorporate in a data analysis,
first because they are based on a fixed set of cosmological parameters, but also because the simulated volume is cubic and periodic.
Each survey group typically needs to run at least one N-Body simulation, 
and measure the power spectrum with and without the measured selection function,
in order to quantify the bias of their measurement. 
The complete approach would then be to run hundred of these to measure the covariance matrix,
and to repeat for a range cosmological parameters values. This whole procedure is expensive, which explains why it is never done in practice. 
The alternative is to use mock galaxy catalogs, obtained, for example, from log normalization of Gaussian densities,
second order perturbation theory (PT), haloPT, and so on.
Unfortunately, the accuracy of such techniques at modeling the four-point functions and angle dependencies has not been fully quantified. 
 
 Another artefact of the simulations is that the number of particles can be arbitrary adjusted such as to 
 suppress the Poisson noise down to a level where it is negligible. This is certainly not true for 
 many galaxy survey, in which the number density is often much lower. We measure a non-Gaussian 
 Poisson error by sampling random fields with a selection threshold chosen as to mimic the number density of a realistic survey,
 and incorporate the effect manually in the analysis, as explained in section \ref{sec:surveys}.

To measure non-Gaussian error bars on a realistic survey, the most accurate procedure
would be to convolve the best available estimator of the  covariance matrix with the selection function.
Because the later is generally not spherically symmetric, it is the full 6-dimensional covariance matrix, $C({\bf k},{\bf k'})$,
that needs to be integrated over.
Let us suppose, for a moment, that we successfully measured that complete non-Gaussian covariance matrix.
It would first contain an element for each Fourier modes ${\bf k}$  (i.e. with no angular averaging),  
and from [Eqs. \ref{eq:P_est} and \ref{eq:cov_th}], we can write:
\begin{eqnarray}
\lefteqn{C_{est}({\bf k}, {\bf k'})  = \frac{\sum_{{\bf k''}, {\bf k'''}} \langle \Delta P({\bf k}'')\Delta P({\bf k}''') \rangle 
|W({\bf k} - {\bf k}'')|^{2} |W({\bf k'} - {\bf k}''')|^{2}}{(N^{2}N_{c}\sum_{{\bf x}}W^{2}({\bf x})w^{2}({\bf x}))^{2}}}  \nonumber \\
\label{eq:non_gauss_cov_est}
\end{eqnarray}
where the angled bracket is nothing else but that full covariance matrix $C({\bf k}'', {\bf k}''')$. 
We can then simplify the result since the covariance between two Fourier modes depends only the angle $\gamma$ between them,
but not on the absolute orientation of the pair in space. 
In other words, we make use of this symmetry argument to write
$C({\bf k}'', {\bf k}''') = C(k'', k''', \gamma)$ without lost of generality.
This angle can further be expressed in terms of the two angles made by ${\bf k}''$ and ${\bf k}'''$ as
\begin{eqnarray}
   \mbox{cos}\gamma = \mbox{cos}\theta''\mbox{cos}\theta''' + \mbox{sin}\theta''\mbox{sin}\theta'''\mbox{cos}(\phi''  - \phi''')
   \label{eq:cos_gamma}
\end{eqnarray}

We show in a later section of this paper that the true covariance matrix can be decomposed
into a sum of factorized terms, each of the form  $F_{1}(k'')F_{2}(k''')G_{1}(\theta'',\phi'')G_{2}(\theta''',\phi''')$.
So the double convolution of [Eq. \ref{eq:non_gauss_cov_est}] can actually be broken
into a sum of smaller pieces, with at most 3-dimensional integrals to perform.

This sets the path of this paper. We propose in the next section a novel technique to 
 measure $C(k'', k''', \gamma)$, we next present the results, and we finally perform the convolution on 
a realistic survey selection functions.




\section{Measuring the Angular Dependence: the Method}
\label{sec:formalism}

As mentioned above, our first objective is to extract the covariance matrix of the power spectra from N-Body simulations, as a function
of two scales and one angle: $C(k,k',\theta)$. 
In this section, 
we develop a novel way to obtain covariances and cross-correlations and which allows us
to perform this measurement.

\subsection{Cross-correlations from Fourier transforms}

We start by assuming we have measured the power spectrum from a large number of simulations.
We first compute the mean of the angle averages:
$\tilde{P}(k) \equiv \langle P({\bf k}) \rangle_{N,\Omega}$ and  the deviation from the mean of each mode:
\begin{eqnarray}
	\Delta P({\bf k}) = P({\bf k})- \tilde{P}(k)
\end{eqnarray}
We then select two scales, $k_i$ and $k_j$, that we want to cross-correlate.
We make two identical copy of three-dimensional power spectra and multiply each one
by a radial top hat function corresponding to the particular scales:
\begin{eqnarray}
 \Delta P_{i}({\bf k}) \equiv \Delta P({\bf k})u_{i}(|{\bf k}|)
\end{eqnarray}
where $u_{i}(k) = \theta(k - k_{i})\theta(-k + k_{i} + \delta k)$ is the product of 
two Heaviside functions. Also, $\delta k$ is the shell thickness, taken to be very small.
We then  cross-correlate the subsets and define: 
\begin{eqnarray}
  \Sigma^{ij}(\Delta {\bf k}) = 
\frac{1}{(2\pi)^{3}}\int \Delta P_{i}({\bf k}) \Delta P_{j}({\bf k}+ \Delta{\bf k})d^{3}k
\label{eq:SigmaCovariance}
\end{eqnarray}

We then express both $\Delta P_{i,j}({\bf k})$'s in [Eq. \ref{eq:SigmaCovariance}]
in terms of their mass auto-correlation functions $\Delta  \xi_{i,j}({\bf x})$.
We first integrate over $\mbox{exp}[i{\bf k} \cdot ({\bf x}+{\bf x'})]d^{3}k$ and obtain a delta function, 
which allows us to get rid of one of the real space integral. After slightly rearranging the terms, we obtain:
\begin{eqnarray}
	\Sigma^{ij}(\Delta {\bf k}) = \int \Delta \xi_{i}({\bf x}) \Delta \xi_{j}^{*}({\bf x}) 
e^{-i\Delta{\bf k}\cdot{\bf x}} d^{3}x
\label{eq:SigmaFFTW}
\end{eqnarray}

\begin{figure}
  \begin{center}
    \centering
    \includegraphics[width=3.0in]{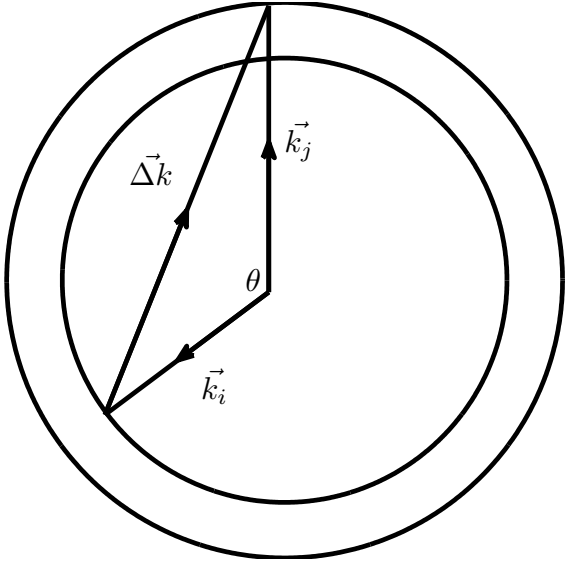}
  \caption{ Geometry of the system. For a fixed pair of shells, the magnitudes of the Fourier modes $k_{i}$ and $k_{j}$ 
  are fixed, so the angle between them is directly found from the separation vector $\Delta {\bf k}$.
Note that we use interchangeably number or roman letters to denote individual Fourier modes.}
    \label{fig:shells}
  \end{center}
\end{figure}

In the above equation, $\Delta \xi_{i}$ can be expressed as:
\begin{eqnarray}
	\Delta \xi_{i}({\bf x}) &= &\frac{1}{(2\pi)^{3}}\int e^{-i{\bf k}\cdot{\bf x}} \Delta P({\bf k}) u_{i}(|{\bf k}|) d^{3}k
	\label{eq:Xi_numerical} \nonumber \\
       & = &\frac{1}{(2\pi)^{3}} \int_{k_{i}}^{k_{i}+\delta k} k^{2} dk \int  e^{-i{\bf k}\cdot{\bf x}}\Delta P({\bf k}) d\Omega
\end{eqnarray}
Since the shells we select are very thin, 
we can safely approximate that the power spectrum is constant over the infinitesimal range,
and thus perform the $k$ integral:
\begin{eqnarray}
     \Delta \xi_{i}({\bf x}) = \frac{1}{(2\pi)^{3}} k^{2}_{i} \delta k \int e^{-i{\bf k_{i}}\cdot{\bf x}}
 \Delta P_{i}({\bf k}) d\Omega
\label{eq:Xi_shell}
\end{eqnarray}
We next repeat the same procedure for a scale $j$, multiply both auto-correlation functions together,
and Fourier transform the product, following [Eq. \ref{eq:SigmaFFTW}].
The result is the cross-correlation $\Sigma^{ij}(\Delta {\bf k})$, 
%
which becomes,  after performing the $x$ integral over the plane wave:
\begin{eqnarray}
    \Sigma^{ij}(\Delta {\bf k})&=& 
    \frac{1}{(2 \pi)^{3}}k_{i}^{2}k_{j}^{2}\delta^{2}k \int d\Omega \int d\Omega' \times\\
     &&\Delta P_{i}({\bf k}) \Delta P_{j}({\bf k'})
    \delta_{D}({\bf k_{j}'} - {\bf k_{i}} -  \Delta{\bf k})
\label{eq:Sigma}
\end{eqnarray}
The delta function enforces $\Delta{\bf k}$ to point from ${\bf k_{i}}$ to ${\bf k_{j}'}$,
which span the shells $k_i$ and $k_j$ respectively in the integrals over the solid angles.
This geometry allows us to use the cosine law and relate $|\Delta{\bf k}|$ to the angle $\theta$ it subtends,
as seen in Fig.  \ref{fig:shells}, such that:
\begin{eqnarray}
	\theta = \mbox{cos}^{-1}\left(\frac{k_{j}^{2} + k_{i}^{2} - |\Delta{\bf k}|^{2}}{2 k_{j} k_{i}} \right)
\label{eq:dk2theta}
\end{eqnarray}
Since many  $\Delta{\bf k}$ subtend the same angle $\theta$, 
we can perform an average over them and compute
\begin{eqnarray}
	\Sigma^{ij}(\theta) \equiv \langle \Sigma^{ij}(\Delta{\bf k})\rangle_{\Delta {\bf k} = \Delta k }  
\label{eq:Sigma_angle_ave}
\end{eqnarray}

\subsection{Normalization}

The quantity $\Sigma^{ij}(\theta)$ is not exactly equal to $C(k_{i}, k_{j}, \theta )$, because there 
is a subtle aliasing effect which is purely geometrical, and which needs to be canceled.
To see how this arises, we work out a very simple scenario, in which the density field is perfectly isotropic.
In that case, we can write $\Delta P({\bf k}) = \Delta P(k)$,
hence the angular integration in [Eq.\ref{eq:Xi_shell}] is straight forward and we get:
\begin{eqnarray}
      \Delta \xi_{i}({\bf x})  = \Delta \xi_{i}(x) = \frac{k^{2}_{i}}{\pi L}   \Delta P_{i}(k) j_{0}(k_{i}x)
\end{eqnarray}
with 
$j_{0}(x)$ the zeroth order spherical Bessel function.
We have also assigned $\delta k = 2\pi/L$ to the shell thickness, which corresponds to the resolution of a simulation of side $L$.
Then, [Eq.\ref{eq:SigmaFFTW}] becomes
\begin{eqnarray}
        \Sigma^{ij}(\theta) = \left(\frac{k_{i}k_{j}}{\pi L}\right)^2 \Delta P(k_{i})\Delta P(k_{j})  F^{ij}(\theta)
\label{eq:norm}
\end{eqnarray}
where
\begin{eqnarray}
    F^{ij}(\theta) 
                 = \int j_{0}(k_{i}x)j_{0}(k_{j}x)j_{0}(\theta x)  x^{2} dx 
\label{eq:F}
\end{eqnarray}
The function $F(k_{i}, k_{j}, \theta)$ is independent of the actual power spectrum;
it is purely a geometrical artifact, and we discuss in Appendix \ref{app:norm} its meaning.
However, a power spectrum that is exactly isotropic should have no angular dependence.
We thus define a normalization $\Sigma^{ij}_{N}(\theta)$,
as the output of [Eq. \ref{eq:Sigma_angle_ave}] with $\Delta P(k_{i,j})=1$ everywhere on the shells.
The final results is obtained by dividing off  this normalization, which cancels off the geometrical effect:
\begin{eqnarray}
    C(k_i,k_j, \theta) \equiv \frac{\Sigma^{ij}(\theta)}{\Sigma^{ij}_{N}(\theta)} =  
\langle \Delta P(k_{i})\Delta P(k_{j}) \rangle
\label{eq:rho_isothropic}
\end{eqnarray}
We stress again that this result is an average over all configurations satisfying
${\bf k_{j}} = {\bf k_{i}} + \Delta{\bf k}$.
To summarize, here is a condensed list of the steps taken to measure $C(k,k', \theta)$:

\begin{enumerate}
\item{Measure the mean angle averaged $\tilde{P}(k)$ from an ensemble of simulations,}
\item{Select a combination of shells $k_{i,j}$ to cross-correlate,}
\item{For each simulation, compute $P({\bf k})$, duplicate and multiply each replica by a top hat $u_{i,j}(k)$,
which effectively sets to zero every off-shell grid cells,}
\item{Subtract $P(k)$ from each cells in the shell,}
\item{Fourier transform both grids, complex multiply them, and Fourier transform back to $k$-space,}
\item{Loop over the $\Delta {\bf k}$ available, bin into $\Sigma(|{\bf \Delta k}|^{2})$, and express the result as a function of $\theta$,}
\item{Repeat steps 5-6, but this time assigning the value of each cell in the shell to unity. Divide $\Sigma(\theta)$ by this normalization.
This is a measure of $C(k_i,k_j, \theta)$ from one simulation,}
\item{Repeat for all simulations, then compute the mean,}
\item{Iterate over steps 2-8 for other shell combinations.}
\end{enumerate}

To achieve better results, we make use of the fact that $P(-{\bf k}) =P({\bf k})$,
hence, following [Eq.\ref{eq:SigmaCovariance}],
we can write $\Sigma^{ij}(-\Delta{\bf k}) = \Sigma^{ij}(\Delta{\bf k})$.
This translates into a theoretical symmetry about $\theta = \pi/2$ in the angular dependence of the covariance.
That property turns out to be very useful at reducing the numerical noise, since we can 
measure the covariance over the full angular range, but fold the results on  to $ 0 < \theta < \pi/2$.
Also, to avoid interpolating error, we chose to bin in $(\Delta k)^{2}$ before transforming to $\theta$. 


\subsection{Zero-lag point}
\label{subsec:zero_lag}

It is important to note that for a given realization, the point at $\theta = 0$, 
which we refer to as the  {\em zero-lag} point, must be treated with care.
We recall from [Eq. \ref{eq:Xi_numerical}] that for a given simulation, 
we subtract  the mean $\langle P(k) \rangle$ from every points which are on the specified $k$-shell
(and the other points are set to zero); we repeat the same procedure on
a second $k$-shell, and we then cross-correlate the two fields. 

When the two shells are identical, i.e. $i=j$, the zero-lag point of each simulation first computes 
the square of the deviation the mean $P(k)$, 
then averages the result over the whole shell.
It is equivalent to calculating the variance over the shell, but
using a mean which is is somewhat off from the actual mean on {\it that} shell. 
That effectively boosts the variance.
When we average over all simulations, the zero-lag points 
can be written as:
\begin{eqnarray}
  \Sigma^{ii}(0) = \langle P_{i}^{2}({\bf k}) \rangle_{N,\Omega} 
  - \langle P({\bf k_{i}}) \rangle_{N,\Omega}^{2}
  \label{eq:rho_ii_zero}
\end{eqnarray}
where, in the first term, the angle average and mean over all realizations are computed {\it after} squaring each grid cell.
By comparison, the variance on angle averaged power spectra would be obtained 
by performing, in the first term, the angle averaging first, then taking the square, then taking the mean.

On the other hand, when the two shells are different, the zero-lag  point is now the average over 
$\Delta P({\bf k})\Delta P({\bf k'})$ on both shells.
Since we are no longer squaring each terms, it now includes  negative terms,
hence is generally of much smaller amplitude.

\section{Validation of the Method}
\label{sec:tests}

We describe in this section a series of tests that compare our numerical results 
to semi-analytical solutions.
We apply our numerical methods on a few simple situations in which we control 
either the density field or the three dimensional power spectrum.
We first test our recipe on a power spectrum that is set to the second Legendre polynomial.
The outcome can be compared to semi-analytical calculations and gives a good 
grip on the precision we can achieve. 
We next measure the angular dependence of the covariance matrix  of white noise densities by 
Poisson sampling many random distributions, and present an estimate for a  non-Gaussian Poisson error
See \citep{2006NewA...11..226C} for discussions of these two types of noise in a cosmological context.
We finally measure the angular cross-correlation from Gaussian random fields
in order to later measure departures from Gaussianity.



\subsection{Testing  $C(k,k',\theta)$ with a Legendre polynomial}
\label{subsec:legendre}

As a first test, we enforce the $z$-dependence of the power spectrum to be equal to the second Legendre polynomial,
and then compare our results to semi-analytic predictions.
We manually set $P({\bf k}) = k_{z}^{2}$, which is thus constant across the $x-y$ plane. 
The mean and the deviation from the mean on a shell are given by 
$\langle P({\bf k}) \rangle_{\Omega} = k^{2}/3$,  
$\Delta P({\bf k}) = (2/3) k^{2}P_{2}(\mu) $ respectively, 
where $P_{\ell}(x)$ is the $\ell$-th Legendre polynomial and $\mu$ is the cosine of the inclination angle.
The mass auto-correlation function associated with this power is
\begin{eqnarray}
  \Delta \xi_{i} (x) = \frac{-2 k_{i}^{4}}{6 \pi L} j_{2}(k_{i}x) 
\end{eqnarray}
The angular dependence of the covariance can be calculated semi-analytically from  Eq.\ref{eq:SigmaFFTW}
and the above mass auto-correlation function.
The angular integration is straight forward, and we obtain
\begin{eqnarray}
    \Sigma^{ij}(\Delta k) = \frac{4 k_{i}^{4} k_{j}^{4}}{9 \pi L} 
    \int_{0}^{\infty} j_{2}(k_{i}x) j_{2}(k_{j}x) j_{0} (\Delta k x) x^{2} dx 
 \end{eqnarray}

\begin{figure}
  \begin{center}
      \centering
    \includegraphics[width=3.2in]{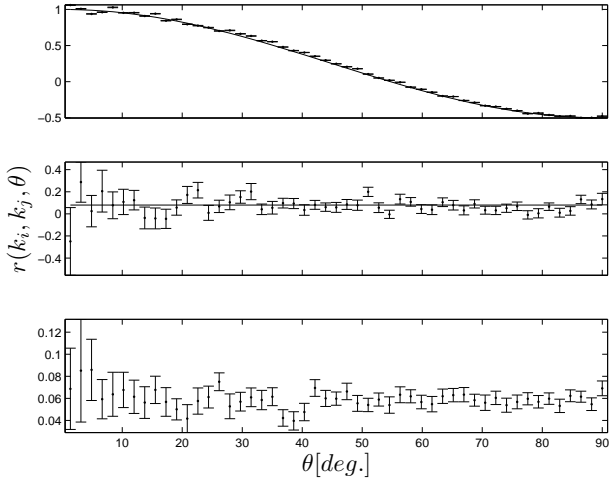}
  \caption{ ({\em  top}:) Angular dependence of the covariance of a power spectrum set to the $2^{nd}$ 
    Legendre polynomial, calculated here at $k_{i=j} =1.0 h\mbox{Mpc}^{-1}$. 
    The solid line is the semi-analytical prediction. The curve is normalized to the value of the zero-lag point,
    such that it represents the actual cross-correlation coefficient between the Fourier modes. In this case,
    modes that points in like-directions are strongly correlated.
    ({\em middle} :) Angular dependence of the power spectrum cross-correlation coefficient 
    measured from $200$ Poisson sampled random fields. 
    The error bars were obtained by $500$ bootstrap sampling.
    We have selected two $k$-shells $i,j$ that are off by one grid cell: $k_{j} = k_{i} + \delta k$,
    with $\delta k  = 0.0314 h/\mbox{Mpc}$ and $k_{i} \sim 1.0 h\mbox{Mpc}^{-1}$. 
    The  distribution for $(i=j)$ is similar in shape, except
    for the zero-lag point which is much larger than any other points, and the plateau that is slightly higher. 
    The solid line in this figure is the predicted value, which is well within the error bars.
    We have reproduced a similar plot for Poisson densities with $8.0$ million peaks, which is 
    also flat, and 
    find that the height of the plateau scales roughly as $1/n^{3}$, 
    where $n$ is the number of Poisson sampled objects.
    ({\em bottom}:) Angular dependence of the power spectrum cross-correlation coefficient, measured from $200$ Gaussian random fields,
    this time with $k_{j} = k_{i} + 5\delta k $ , and again $k_{i} \sim 1.0 h\mbox{Mpc}^{-1}$. 
    The theoretical prediction is zero, whereas we measure a constant $6$ per cent correlation bias across all angles. 
    We have verified that this bias is scale independent by changing $k_{i,j}$.}
    \label{fig:test_cov}
  \end{center}
\end{figure}

We perform the $x$ integral with $k_{i=j}=1.0 h\mbox{Mpc}^{-1}$, repeat the procedure for $\Sigma^{ij}_{N}(\Delta k)$, 
and obtain a semi-analytical prediction: $C(k,k',\theta) \sim P_{2}(\mbox{cos}\theta)$, up to numerical noise.
This agrees well  with  the numerical results produced by our technique, as shown in the top part of Fig.  \ref{fig:test_cov}.
We are plotting the angle-dependence of the covariance matrix, normalized by the angle average of the covariance,
such that the curve  represents the actual cross-correlation coefficient between the Fourier modes.
We mention here that in the case where $k_{i} \neq k_{j}$, which we encounter in the following sections, 
we normalize to the square root of the product of the corresponding matrix elements:
\begin{eqnarray}
r(k_i,k_j,\theta) = \frac{C(k_i,k_j,\theta)}{\sqrt{C(k_i,k_i)C(k_j,k_j)}}
\label{eq:crosscorr}
\end{eqnarray}
In the particular case under study in this section, the Fourier modes separated by small angles are strongly correlated 
by construction.


\subsection{Testing $C(k,k',\theta)$  with Poisson-sampled random fields}
\label{subsec:poisson}

To measure the response of our code to white noise, we produce a set of $200$ density field representing
Poisson sampling of random distributions. 
The sensitivity threshold is chosen such that $\sim8000$ peaks were counted on average. 
The standard deviation in the measured $P(k)$ 
decreases roughly as $k^{-2}$, expected from the fact that the number of cells on a $k$-shell grows as $k^{2}$.

\begin{figure}
  \begin{center}
      \centering
    \includegraphics[width=3.2in]{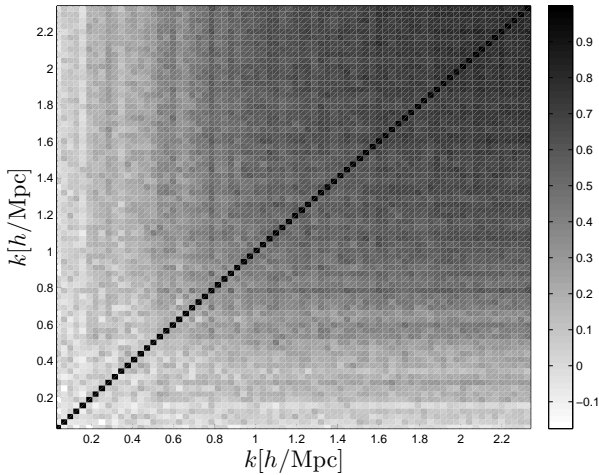}
  \caption{ Cross-correlation coefficient matrix, 
    measured from the power spectra of 200 Poisson sampled random densities fields, selected to have $8000$ peaks on average.
    The correlation in high $k$-modes is unphysical, as explained in the text, and this represents our estimate of the non-Gaussian Poisson uncertainty.}
    \label{fig:poisson_cross_corr}
  \end{center}
\end{figure}

Because of the random nature of Poisson density,  
the variance on a given shell should be roughly constant across all directions.
Moreover, after averaging  over many realization, Poisson densities are in principle statistically isotropic.
We thus expect the measured angular dependence of the covariance to be very close to flat,
and, from [Eq.{\ref{eq:rho_isothropic}}], we estimate it should plateau at a value somewhat
similar to the covariance of angle averaged power spectrum, $C(k,k')$:
\begin{eqnarray}
C_{Poisson}(k,k',\mu) \sim C_{Poisson}(k,k') + A\delta_{kk'}\delta_{\mu \pm 1}
\label{eq:cov_Poisson}
\end{eqnarray} 
where $\mu = \mbox{cos}\theta$ and the two delta functions
ensure that modes with different directions or scales do not couple together.
Also, $C_{Poisson}(k,k')$ is obtained directly from the covariance matrix of the angle average power spectra.
The constant $A$ is much larger than  $C_{Poisson}(k,k')$, for reasons explained in section \ref{subsec:zero_lag},
but the precise value is irrelevant to the current analysis. 
Fig.   \ref{fig:poisson_cross_corr} shows the cross-correlation coefficient matrix for non-Gaussian Poisson noise.
We observed that  the angle-averaged modes are correlated by  more than $30$ per cent between scales smaller than $ k =1.0 h/\mbox{box}$.
The reason for this feature is actually independent of cosmology,
even though the matrix has a look very similar to that measured from simulations\footnote{It is in fact arguable that such a matrix, constructed from a set of Poisson densities, could have better performances at modeling the `true' non-Gaussian covariance matrix, compared to the naive Gaussian approximation.} .
The explanation lies in the fact that our Poisson densities 
do not have {\it exactly} the same number of objects, 
hence the asymptotic value of $P(k)$ is not a perfect match for all field. 
This slight scatter in power translates into a correlation between the high $k-$modes 
of a {\it given} density field, and we use this matrix to estimate the non-Gaussian Poisson noise.
This is in good agreement with the predictions of \citep{2006NewA...11..226C},
which calculated that the Poisson sampling of Gaussian  fields induce non-Gaussian statistics,
and that well separated scales can correlate significantly.


We then measure the angular dependence of the covariance for these $200$ Poisson distributions, also at $k \sim 1.0 h\mbox{Mpc}^{-1}$.
We obtain a distribution which is indeed close to flat, and consistent with a uniform $10$ per cent correlation, as shown in 
the middle plot of Fig.  \ref{fig:test_cov}.
As before, we have normalized the plot by the angle averaged covariance matrix element, 
such as to exhibit the angular cross-correlation.
Because the zero-lag point is typically a few orders of magnitude above the other points,
we quote its value in the text or in the figures' caption where relevant, and resolve the structure of the other angles. 
The mean of the un-normalized distribution is $133.3 \mbox{Mpc}^{6}h^{-6}$, a $10$ per cent agreement
with our rough estimate. 
We have re-binned the distributions on to a set of points
that are optimal for the upcoming angular integration, as described in section \ref{sec:Cl}.  

\subsection{Testing $C(k,k',\theta)$ with Gaussian random fields}
\label{subsec:test_gauss}

The next test consists in measuring the angular dependence of the covariance from of 200 Gaussian random fields.
We use $200$ power spectra measured at $z = 0.5$, obtained from N-Body simulations (section \ref{subsec:nbody}),
to generate 200 fields. 
Similarly to the Poisson fields, we expect the distribution to be be overall flat,
except for the zero-lag point.
Because we choose not to Poisson sample these Gaussian densities,
the randomness should be such that near to perfect cancellation occurs between 
the different angles, and the plateau should be at zero.
In the continuous case, the Gaussian covariance can be expressed as 
[Eq. \ref{eq:Cl_rho}], with 
\begin{eqnarray}
C_{Gauss}(k_i, k_j, \mu) = \frac{2 \langle P(k_i) \rangle^{2}}{N(k_i)}\delta_{ij}\delta_{\mu, \pm 1}
\label{eq:rho_gauss}
\end{eqnarray}
where $N(k)$ is the number of Fourier modes in the $k$-shell.
For $k_i=k_j$, the  zero-lag point contains perfectly correlated power,  
so  we expect  it to have a very large value. 
As explained in section  \ref{subsec:zero_lag}, we cannot directly compare its value
to $2P^{2}(k)/N(k)$, since the former is bin dependent, while the latter is not.
In the case where  $i  \ne  j$ however, the zero-lag point should drop down to numerical noise.

The measured angular dependence is presented in the bottom part of Fig.  \ref{fig:test_cov}, where we see that 
the distribution is flat and consistent with $6$ per cent correlation. 
This indicates that our method suffers from a small systematic bias and detects
a small amount of correlation, in a angle independent manner. 
We have repeated this measurement for different scales $k_{i,j}$ and obtained the same bias.
We therefore conclude that any signal which is smaller than this amount is bias dominated and 
not well resolved.

\section{Measuring the Angular Dependence}
\label{sec:rho}

In this section, we present the results obtained from the measurement of 
the angular covariance in our $200$ simulations.
We explore different scale combinations and attempt to compare the outcome to expected results whenever possible.
In particular, the linear regime should somewhat reproduce the behavior we observe in Gaussian random fields 
(see section \ref{subsec:test_gauss}).
In all figures, the error bars were obtained from $500$ bootstrap re-sampling of our simulations, unless otherwise specified.


\subsection{N-body simulations}
\label{subsec:nbody}

Since our Universe is not Gaussian at all scales relevant for BAO or weak lensing analyses, 
a robust  error analysis should be based on non-Gaussian statistics,
and, as mentioned earlier, N-body simulations are well suited to measure covariance matrices.
Our numerical method is fast enough that, for fixed $k_i$ and $k_j$, 
we can compute the angular dependence of the covariance matrix in about one minute. 
The average over 200 realizations can be done in parallel, hence producing all available 
combinations takes very little time.

\begin{figure}
  \begin{center}
\centering
\includegraphics[width=3.2in]{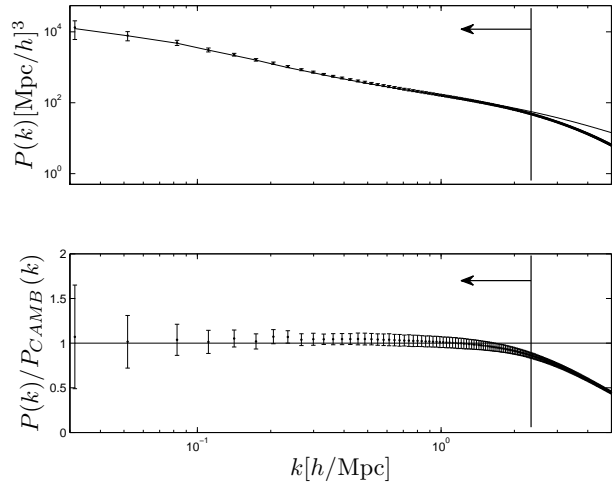}
  \caption{({\it top:}) Power spectrum of 200 simulations, produced by {\tt cubep3m}, 
    compared to {\tt CAMB} at $z=0.5$ (solid line).
    The error bars are the $1\sigma$ standard deviation on the $200$ measured $P(k)$.
    We only include modes with $k \le 2.34 h\mbox{Mpc}^{-1}$ in this analysis, as indicated by the arrow in the figure.
    ({\it bottom:}) Ratio between the simulated and modeled power spectra.}
    \label{fig:sim_power}
  \end{center}
\end{figure}

The simulations are produced by {\tt cubep3m}\citep{2005NewA...10..393M}, 
a public N-body code that is both {\tt openmp} and 
{\tt mpi} parallel, which makes it among the fastest on the market\footnote{http://www.cita.utoronto.ca/mediawiki/index.php/CubePM}. 
We generate 200 Gaussian distributions
of 200 Mpc$h^{-1}$ per side, with $256^{3}$ particles, starting  at $z_{i}=40$, and evolved them until $z=0.5$.
The simulations are run on the CITA Sunnyvale cluster, a Beowulf cluster of 200 Dell PE1950 compute nodes,
each equipped with 2 quad cores Intel(R) Xeon(R) E5310 @ 1.60GHz processors. 
Each node has access to 4GB of RAM and 2 gigE network interfaces.
The power spectrum  of these simulations is shown in Fig.  \ref{fig:sim_power}, and shows
a good agreement with the non-linear predictions from {\tt CAMB} \citep{1996ApJ...469..437S},
up to $k \sim 0.25 h\mbox{Mpc}^{-1}$.  Beyond that scales the structures are under estimated 
due to the resolution limit of the simulations.
For the rest of this paper, we only consider well resolved scales, in occurrence
those in the range $k \in [0.314, 2.34] h\mbox{Mpc}^{-1}$, which we organize into $75$ linearly spaced bins.

\subsection{Results}
\label{subsec:results_rho}

\begin{figure}
  \begin{center}
  \centering
   \includegraphics[width=3.2in]{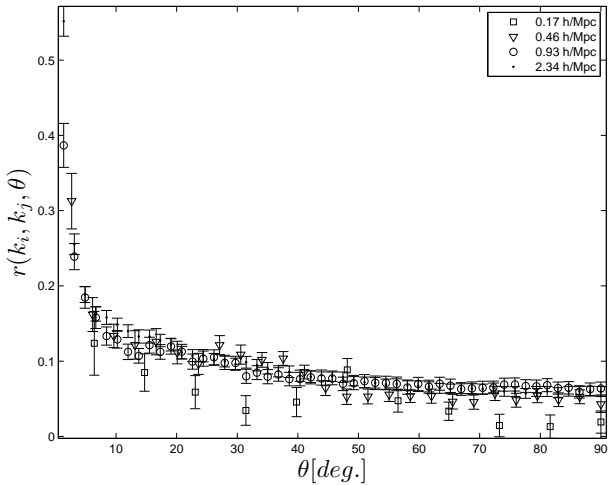}
   \caption{ Angular dependence of the power spectrum cross-correlation, 
    measured from  of 200 density fields, at $k_{i=j} = 0.17,  0.46, 0.93$ and $2.34 h \mbox{Mpc}^{-1}$. 
    The distribution exhibits a correlation of less than $10$ per cent for angles
    larger than about $30^{o}$. For scales less than $0.5h\mbox{Mpc}^{-1}$, the correlation increases up to $15$ per cent
    for angles smaller than $10^{o}$, and to more than $40$ per cent for smaller angle.}
    \label{fig:sim_cov_on_diag}
  \end{center}
\end{figure}

\begin{figure}
  \begin{center}
  \centering
   \includegraphics[width=3.2in]{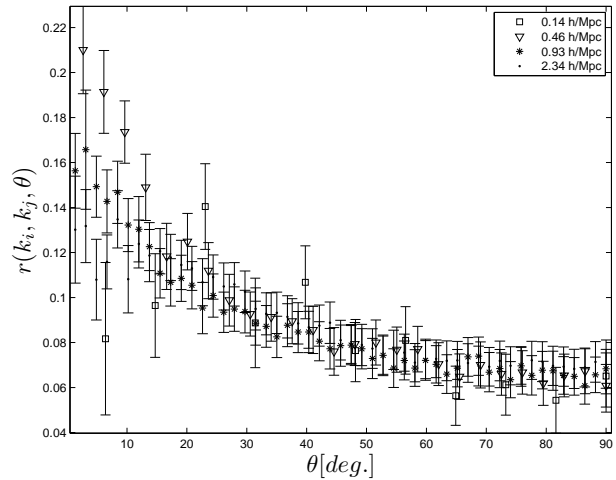}
    \caption{ Angular dependence of the power spectrum cross-correlation, 
    measured from  of 200 density fields, at $k_{i} = 0.61$, and $ k_{j} = 0.14,  0.46, 0.93$ and $2.34 h \mbox{Mpc}^{-1}$. 
    The distribution exhibits a correlation of less than $10$ per cent for angles
    larger than about $30^{o}$. For scales of similar sizes, the correlation increases up to $15-20$ per cent
    for angles smaller than $15^{o}$.}
    \label{fig:sim_cov_off_diag}
  \end{center}
\end{figure}

We present in Figs. \ref{fig:sim_cov_on_diag} and \ref{fig:sim_cov_off_diag} the angular dependence of the covariance 
between the power spectrum of various scales. As explained in the previous section, the distributions are normalized 
such as to represent the cross-correlation coefficient between modes separated by an angle $\theta$.
In the first figure, both scales are selected to be identical, and vary 
progressively from $k = 0.17 h\mbox{Mpc}^{-1}$ to $2.34 h\mbox{Mpc}^{-1}$. 
Modes separated by an angle larger than $30^{o}$ are less correlated at all scales, 
and the correlation is even smaller for modes smaller than $0.5h\mbox{Mpc}^{-1}$.
These latter modes are grouped in larger bins due to the higher discretization of the shells,
and ideally one would like to run another set of simulation at larger scales
to have a better resolution on those scales. However, the modeling of these rather larger scales 
have very little impact on the non-Gaussian analysis we are carrying, we therefore
do not attempt to improve the situation.
For highly non-linear scales, the correlation between modes less than $10^{o}$ increases up to  $55$ per cent.
 
In the second figure, one of the two scale is held constant,
at $k = 0.61 h\mbox{Mpc}^{-1}$, while the other varies over the same range.
Modes separated by angles larger than $30^{o}$ are less than $10$ per cent correlated, 
for all combinations of scales. When the two scales are of comparable size, the 
the correlation climbs up to values between $15$ and $20$ per cent for angles smaller than $15^{o}$.   
 
This angular behavior is enlightening, as it shows how the error between Fourier modes separated only by a small angle tend to correlate first.
Qualitatively, this validate the fact that in non-Gaussian densities, 
quasi-parallel Fourier modes are probing  essentially the same collapsed structures. 
When the angle is closer to $90^{o}$, however, one mode could go along a filament and the other across it, 
producing only  weak correlations. It could thus be possible to construct a highly clustered density in which we could observe an anti-correlation at $90^{o}$,
provided we are not noise dominated.



This coherent behavior is a clear sign that the non-linear structure underwent gravitational collapse,
and the departure from Gaussianity and white noise is obvious. 
Another signature of non-Gaussianity is that even in the presence of a small offset between the scales,
the small angle correlation has a value higher than those at larger angles, 
because of the coupling between those scales. 
Fig.  \ref{fig:sim_cov_off_diag} shows this effect. 

\subsection{From $C(k_i,k_j,\theta)$ to $C(k_i,k_j)$}
\label{sec:rho_to_cov}

It is possible to recover the covariance matrix one obtains
from the angle averaged $P(k)$  by performing  a weighted sum 
over the angular covariance\footnote{The weight here is simply the number of contribution that enter each angular bin, 
divided by the square of the total number of cells on the $k$-shell. 
In other words, because the angular covariance we measure is an average over many pairs of cells,
that average must first be undone, then the different angles are summed up, and we finally divide
by the total number of contributions.}.
Another test of the accuracy of our method  is thus to compare the two ways that measure $C(k_i,k_j)$. 
This is by far the least convenient way of measuring  this matrix, and we perform this check
solely for verification purposes.

\begin{figure}
  \begin{center}
   \centering
   \includegraphics[width=3.2in]{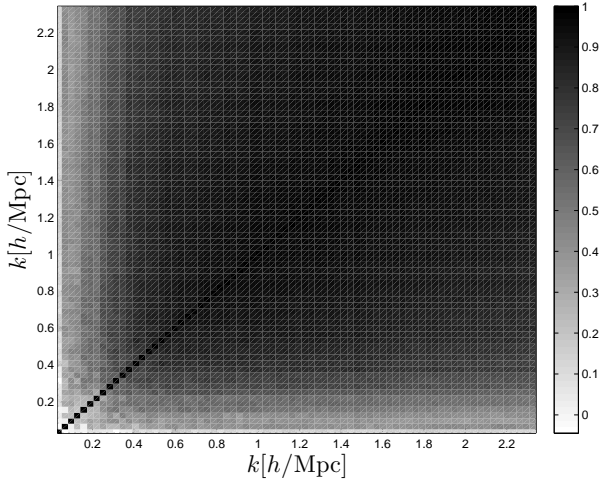}
  \caption{ Cross-correlation coefficient matrix, as measured from the angular covariance. 
    Each matrix element $i,j$ was obtained from a reweigthed sum over $C(k_i,k_j,\theta)$. 
    This is consistent with matrices previously measured in the literature \citep{2005MNRAS.360L..82R,2009ApJ...700..479T, 2011arXiv1106.5548N}}
    \label{fig:rho_cross_corr}
  \end{center}
\end{figure}

\begin{figure}
  \begin{center}
    \centering
   \includegraphics[width=3.2in]{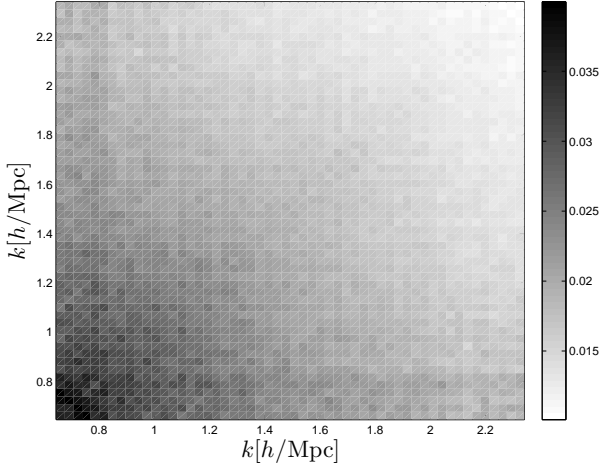}
  \caption{ Fractional error between the covariance matrices obtained with the two methods.
    We have suppressed the largest scales, which are noisy due to low statistics, 
    and present the percent level agreement at smaller scales.}
    \label{fig:cross_corr_frac_error}
  \end{center}
\end{figure}

We perform this weighted sum and construct $C(k_i,k_j)$, then 
compute a similar matrix from our $200$ angle averaged power spectra.
We present in Fig.  \ref{fig:rho_cross_corr} the cross-correlation coefficient matrix (see [Eq. \ref{eq:crosscorr}])
obtained in the first way, and show the fractional error between both methods in  Fig.   \ref{fig:cross_corr_frac_error}.
%
%
We observe that they agree at the few percent level, so long as we are in the non-linear regime.
At very low $k$-modes, however, many matrix elements are noisy due to the discretization of the shell; 
the $(\Delta k,\theta)$ mapping in this very coarse 
grid environment becomes unreliable, and the re-weighting hard to do correctly.
This results in high fractional errors, but at the same time,  this region
is still in the regime where the analytic Gaussian prediction is valid.
%
In addition,  this paper attempts to solve the bias caused by the non-Gaussianities that lie in the trans-linear and non-linear regime,
in which discretization effects are much smaller. 
Finally, we recall that these matrix elements have very little impact on most parameter studies since such scales 
contain almost no Fisher information \citep{2005MNRAS.360L..82R, 2011arXiv1106.5548N}.


\section{Multipole Decomposition}
\label{sec:Cl}

As shown in last section,  we have extracted the covariance matrix $C(k_i, k_j, \theta)$ of matter power spectrum,
cross-correlating the $75$ different scales selected. 
Since the final objective is to incorporate this massive object into generic data analysis pipelines, 
it must be somehow simplified or made more compact. A quick glance at the figures of section \ref{sec:rho}
reveals that the angular dependence of the covariance can be decomposed into a series  of Legendre polynomials,
in which only a few multipoles will bear a significant contribution.   
This allows us to rank the multipoles by importance and to keep only the dominant ones. 
These results are further simplified in section  \ref{sec:factorization}, where we provide fitting formulas
to reconstruct  $C(k_i, k_j, \theta)$.

In this section, we describe how we perform this spherical harmonic decomposition,
then we test our method on the control samples described in section \ref{sec:tests}, 
and we finally measure the $C_{\ell}(k,k')$  from the simulations.

\subsection{From $C(k_{i},k_{j},\theta)$ to $C_{\ell}(k_{i},k_{j})$ }

Here we lay down the mathematical relation between  $C(k_i, k_j, \theta)$ and $C_{\ell}(k_{i},k_{j})$.
Let us first recall that the spherical harmonics $Y^{\ell m}(\theta,\phi)$ 
can serve to project any function $F(\theta, \phi)$ on to a set of $a_{\ell m}$ as:
\begin{eqnarray}
  a_{\ell m} = \int Y^{\ell m}(\theta,\phi) F(\theta, \phi) d\Omega
\end{eqnarray}
We substitute $F(\Omega) \rightarrow \Delta P_{i}({\bf k}) = \Delta P_{i}(k,\Omega)$,
which causes the coefficients to be scale dependent, i.e.  $a_{\ell m}  \rightarrow a_{\ell m}(k)$.
The angular power spectrum at a given angular size $\theta \sim 1/\ell$ is  defined as
\begin{eqnarray}
  C_{\ell}(k_{i},k_{j}) \equiv \frac{1}{2\ell + 1}\sum_{m=-\ell}^{\ell}|a_{\ell m}(k_{i})a_{\ell m}^{*}(k_{j})|
\end{eqnarray}
Combining both equations, and writing $C_{\ell}^{ij} \equiv C_{\ell}(k_{i},k_{j})$ to clarify the notation,  we get
\begin{eqnarray}
  C_{\ell}^{ij} &=& \lefteqn{\frac{1}{2\ell + 1}\sum_{m=-\ell}^{\ell}\int Y^{\ell m*}(\Omega')Y^{\ell m}(\Omega) \times} \nonumber\\
 && \Delta P(k_{i}, \Omega) \Delta P^{*}(k_{j}, \Omega') d\Omega d\Omega'
\end{eqnarray}
We  use the completion rule on spherical harmonics to perform the sum:
\begin{eqnarray}
\sum_{m=-\ell}^{\ell} Y^{\ell m}(\Omega)Y^{\ell m}(\Omega') = \frac{2\ell+1}{4\pi}P_{\ell}(\mbox{cos} \gamma)
\label{eq:Y_lm_sum}
\end{eqnarray}
where $\gamma$ is the angle between the $\Omega$ and $\Omega'$ directions, 
and where $P_{\ell}(x)$ are the Legendre polynomials of degree $\ell$.
We then write
\begin{eqnarray}
   C_{\ell}^{ij} = \frac{1}{4\pi}\int \Delta  P(k_{i}, \Omega) \Delta P^{*}(k_{j}, \Omega') P_{\ell}(\mbox{cos} \gamma) d\Omega d\Omega'
\end{eqnarray}
Since we know that ${\bf k_{i}} + \Delta{\bf k} = {\bf k_{j}}$, 
we make a change of variable and rotate the prime coordinate system such that ${\bf k}$ always
points towards the $z$-axis. In this new frame, we have $d\Omega'' = d\mbox{cos}\theta'' d\phi''$,
where $\theta''$ is the angle subtended by $\Delta{\bf k}$. 
$\theta''$ thus corresponds to the angle between the two Fourier modes ${\bf k}$ and ${\bf k'}$.
It is also equal to $\gamma$ in [Eq. \ref{eq:Y_lm_sum}].
We perform the `unprime' integral first, which gives
\begin{eqnarray}
 C_{\ell}^{ij} = \frac{1}{4\pi}\int P_{\ell}(\mbox{cos} \gamma) 
 \int \Delta P_{i}({\bf k})\Delta P_{j}({\bf k} + \Delta{\bf k}) d\Omega d\Omega''  
\label{eq:C_ell_integral}
\end{eqnarray}
The inner integral is $C(k_i,k_j,\gamma)$, we rename $\gamma \rightarrow \theta$ and obtain
\begin{eqnarray}
C_{\ell}^{ij} = \int P_{\ell}(\mbox{cos}  \theta) C(k_i,k_j,\theta)d\Omega
\label{eq:Cl_rho}
\end{eqnarray}


In practice we are dealing with a discretized grid, hence we must convert the integral of [Eq.\ref{eq:Cl_rho}] into a sum.
To minimize the error, we use a Legendre-Gauss weighted sum, the details of which can be found in the Appendix.
In order to validate our method, we designed a few tests which are explained in the following sections.


\subsection{Testing $C_{\ell}$ with a Legendre polynomial, with Poisson and Gaussian distributions}

We start our tests by measuring the $C_{\ell}(k_i,k_j)$ from the angular dependence of the covariance of power spectra,
which is explicitely set to the second Legendre polynomial on the selected $k$-shells, 
as described in section \ref{subsec:legendre}. 
We expect the projection to produce a delta function 
at $\ell = 2$, up to numerical precision, since the Legendre polynomials are mutually orthogonal.
%
We observe from this simple test a sharp peak at $\ell = 2$, 
which is about two orders of magnitude higher than any other points.

\begin{figure}
  \begin{center}
\centering
\includegraphics[width=3.2in]{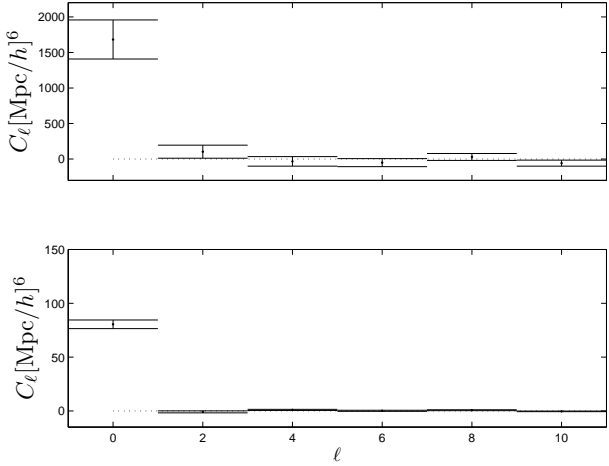}
  \caption{ 
    ({\em top} :) Angular power of the cross-correlation obtained from 200 Poisson densities, at $k_{i \sim j} \sim 1.0 h\mbox{Mpc}^{-1}$,
    with an offset of one grid cell  between the two scales, corresponding to $\delta k = 0.0314 h \mbox{Mpc}^{-1}$. 
    The power at $\ell \ne 0$ is consistent with zero, as expected from [Eq. \ref{eq:Cl_Poisson}].
    We recall that the angular dependence of the covariance from Poisson densities is very weak,
    hence it projects almost exclusively on the $\ell=0$ term.  
    ({\em bottom} :)  Gaussian angular power at $k_{i} \sim 1.0 h\mbox{Mpc}^{-1}$, and $k_{j}  =  k_{i}  + 5\delta k$.
  		The analytical prediction is zero at all multipole,  while we measure a $C_{0}$ term of about  $80.5 h^{-6} \mbox{Mpc}^{6}$. 
  		This is caused by the $6$ percent bias we observed in  Fig.  \ref{fig:test_cov}. }
    \label{fig:Cl_test}
  \end{center}
\end{figure}


We next measure the $C_{\ell}$ from the covariance matrix of Poisson densities, 
whose angular dependence, we recall, is close to flat (see section \ref{subsec:poisson}),
up to a delta function on the zero-lag point when the two shells are identical. 
From the orthogonality of the Legendre polynomials, 
a flat distribution is projected exlusively on the first multipole,
we thus expect $C_{\ell}^{Poisson}(k \neq k')$ to peak at $\ell=0$,
and to vanish for other $\ell$. 
Moreover, we expect the $C_{\ell}^{Poisson}(k = k')$ to exhibit,  in addition, a vertical shift caused by the integration over the zero-lag point.
The analytical expression can be obtained from [Eqs. \ref{eq:cov_Poisson},\ref{eq:Cl_rho}].
The azimutal integration gives a factor of $2\pi$, the $\mu$ delta function gets rid of the last integral, and we get:
\begin{eqnarray}
C_{\ell}^{Poisson}(k,k') =  2 \pi C_{Poisson} \frac{2}{2\ell + 1} \delta_{\ell 0}                       & , k \ne k' \nonumber \\
C_{\ell}^{Poisson}(k,k') =  2 \pi C_{Poisson} \frac{2}{2\ell + 1} \delta_{\ell 0} + 4 \pi A\delta_{kk'} & , k = k'
\label{eq:Cl_Poisson}
\end{eqnarray}
The only scale dependence thus comes from the surface of the $k$-shell, and thus drops as 
$k^{-2}$, as explained in section \ref{subsec:poisson}.

In the $k \neq k'$ case, we find that in the non-linear regime, the $\ell=0$ point is at least two orders of 
magnitude above the other even $\ell$, and 18 orders above the odd $\ell$.
The results are presented in the middle part of Fig.  
\ref{fig:Cl_test} for $k_{i \sim j} \sim 1.0 h \mbox{Mpc}^{-1}$.
The error bars were obtained from a bootstrap resampling of the angular dependence 
of the covariance matrix, measured from 200 Poisson densities.
When  $k = k'$, we find that the zero-lag point effectively shifts the whole distribution upwards
by an amount equivalent to $4\pi C^{Poisson}(k,k,0)$.

Finally, we compare the $C_{\ell}$ distribution measured from Gaussian fields  to an analytical prediction,
obtained from [Eqs. \ref{eq:Cl_rho},\ref{eq:rho_gauss}].
\begin{eqnarray}
C_{\ell}^{Gauss}(k,k') = 2\pi \frac{2 \langle P(k) \rangle^{2}}{N(k)} (1+(-1)^{\ell})\delta_{kk'}
\label{eq:Cl_Gauss}
\end{eqnarray}
which is null for odd-$\ell$.

We measured the Gaussian $C_{\ell}$ from the covariance matrix of 200 Gaussian random fields,
as outlined in section \ref{subsec:test_gauss}. 
We show the results in the bottom part  of Fig.  
\ref{fig:Cl_test} for the case where there is a slight offset between the two scales, in which case the analytical prediction is zero.
Our results are consistent with  zero for all multipoles except  $\ell=0$, which receives an undesired contribution 
from the constant $6$ per cent bias described in section \ref{subsec:test_gauss}
and observed in Fig. \ref{fig:test_cov}.
It turns out that this $C_{0}$ contribution is very small compared (i.e. less than one per cent) to the values 
obtained from simulated density fields, hence we do not attempt to correct for it.
In the case where the two shells are identical,  we observe similar results, up to an upward shift caused by the zero-lag point,
which propagates to all multipoles.

When performing this decomposition on the covariance matrix obtained from actual density fields, 
the departure from Gaussianity can be quantified
as the number of distinct $\ell$ needed to describe the angular power distribution.


\subsection{Measuring $C_{\ell}(k_{i},k_{j})$ from simulations}
\label{subsec:C_ell_den}

We now present in this section the multipole 
decomposition of the $C(k_i,k_j,\theta)$ matrix measured from our simulations. 

\begin{figure}
  \begin{center}
\centering
\includegraphics[width=3.2in]{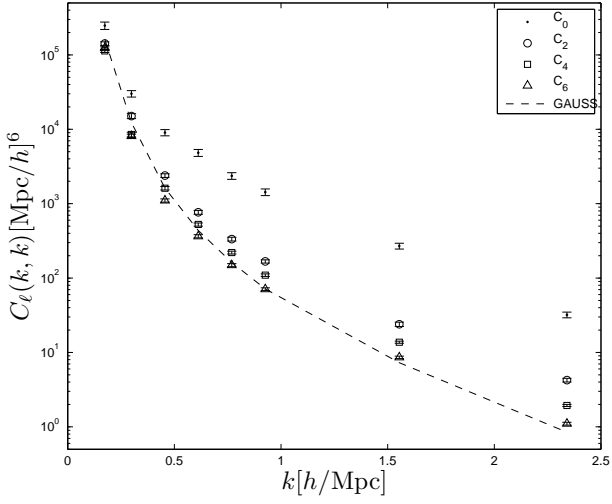}
  \caption{ Angular power of 200 densities, where $k_{i=j}$. 
    The dashed line is the Gaussian prediction, obtained from [Eq. \ref{eq:Cl_Gauss}].
    From this figure, we observe that the diagonal of multipoles higher than $\ell=4$ converge to the Gaussian predictions.}
    \label{fig:Cl_sim_ondiag}
  \end{center}
\end{figure}

We show in Fig. \ref{fig:Cl_sim_ondiag} the first few non-vanishing multipole moments (i.e. $\ell = 0,2,4,6$),
in the case where both scales are exactly equal. We observe that higher multipoles become closer to the Gaussian prediction 
given by [Eq.  \ref{eq:Cl_Gauss}], and in fact only the first three differ enough to have a non-negligible impact. 
All the error bars in this section were obtained from bootstrap re-sampling.
As we progress deeper in the non-linear regime, we expect to encounter a mixture of the following two effects:
an increase in the number of $\ell$  required to specify the $C_{\ell}$ distribution, 
or in the departure from the Gaussian predictions of the $C_{\ell}$ at fixed $\ell$.
As seen from Fig.  \ref{fig:Cl_sim_ondiag}, the departure between the multipoles and the Gaussian
power increases for higher $k$-modes,  an effect prominent in the first multipole.
The departure becomes more modest for higher multipoles, and eventually 
we cannot distinguish between Gaussian and non-Gaussian.
This suggests that the non-Gaussianities are encapsulated in the second of the effect above mentioned.

\begin{figure}
   \begin{center}
   \centering
\includegraphics[width=3.2in]{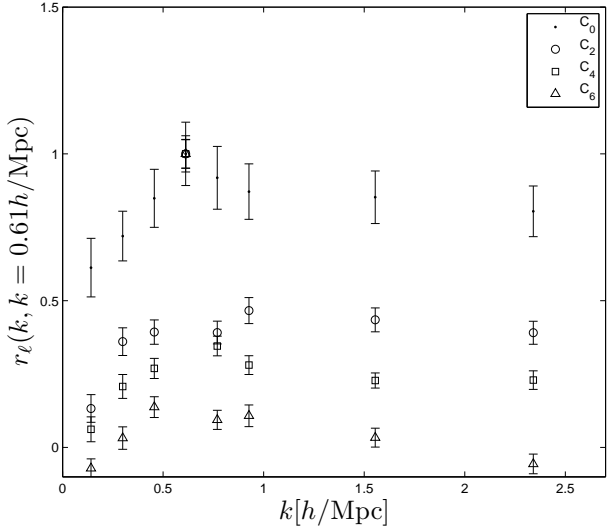}
    \caption{ Same as Fig.  \ref{fig:Cl_sim_ondiag}, but with $k_{i} = 0.61 h \mbox{Mpc}^{-1}$ being held.
     The Gaussian prediction is zero in this case. The measurements are normalized by the square root of their diagonal contributions,
     such as to show the relative importance of each multipole. As $\ell$ increases, the off-diagonal contribution  becomes smaller, even for
     combinations of scales similar in amplitudes.  The fourth point starting from the left is identical to unity for all multipoles, 
     as it corresponds to a diagonal point.}
    \label{fig:Cl_sim_offdiag}
  \end{center}
\end{figure}

We then show in Fig.  \ref{fig:Cl_sim_offdiag} the same multipole moments, 
this  time for the case where one scale is fixed at $k=0.61 h \mbox{Mpc}^{-1}$, while the other is allowed to vary. 
Once again,  higher multipoles have smaller amplitudes, and approach the Gaussian prediction off zero.  



On the diagonal, the relative difference between the multipoles in the linear regime  becomes
smaller and converge to the predicted value, as expected. 
In addition, in the linear regime, the angular  power of the off-diagonal elements (i.e. $k_i \ne  k_j$) is one to two 
orders of magnitude smaller than the diagonal counter part. 
As we progress to the non-linear regime however, the off-diagonal elements
decrease less rapidly,  and a convenient way to express this is to look at the cross-correlation coefficient matrices.





\subsection{$C_{\ell}(k,k')$ matrices}
\label{subsec:ClMatrix}

As mentioned above, we need to look at an individual $\ell$, for all scales combinations $(k,k')$,
and find the multipole beyond which the off-diagonal elements become negligible.
The whole purpose behind this is to model the full covariance matrix as:
\begin{eqnarray}
   C(k,k',\theta) =  \frac{1}{4\pi}\sum_{\ell  =0}^{\infty}  (2\ell + 1) C_{\ell}(k,k') P_{\ell}(\mbox{cos}\theta)
\label{eq:cov_expansion}
\end{eqnarray} 
where the lower $\ell$ terms are measured from our simulations, and the others obtained from 
the Gaussian analytical prediction ([Eq.\ref{eq:Cl_Gauss}]).

\begin{figure}
  \begin{center}
   \centering
\includegraphics[width=3.2in]{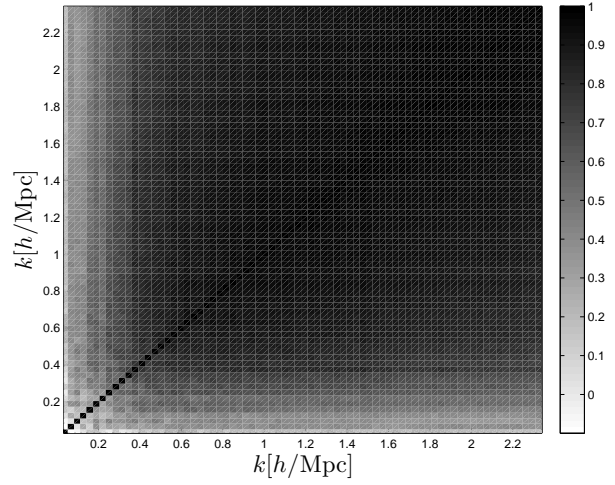}
  \caption{$C_{0}$ matrix, normalized such that the diagonal elements are equal to unity.
    This matrix is completely equivalent cross-correlation coefficient matrix of angle averaged $P(k)$, 
    up to a factor of $4\pi$. It represents the correlation between different scales, and shows that 
    scales smaller than $k \sim 1.0 h\mbox{Mpc}^{-1}$ are correlated by more than $80$ per cent.
    }
    \label{fig:C0_matrix}
  \end{center}
\end{figure}

In the figures of this section, we present these  `$C_{\ell}$' matrices, 
normalized to unity on the diagonal. 
These are thus in some sense equivalent to cross-correlation coefficient matrices.
Fig.  \ref{fig:C0_matrix} presents the normalized $C_{0}$ matrix, which shows a structure similar to that of the 
cross-correlation coefficient matrices obtained previously (Fig.  \ref{fig:rho_cross_corr}).
The resemblance is not surprising, since $C_{0}  = 4\pi C(k,k')$,
(we can see this by setting $P_{\ell = 0} = 1$ in [Eq. \ref{eq:C_ell_integral}]) 
The $C_{0}(k,k')$ matrix thus contains
the information about the error bars of angle averaged power spectra, 
as well as their correlation.

\begin{figure}
  \begin{center}
\centering
\includegraphics[width=3.2in]{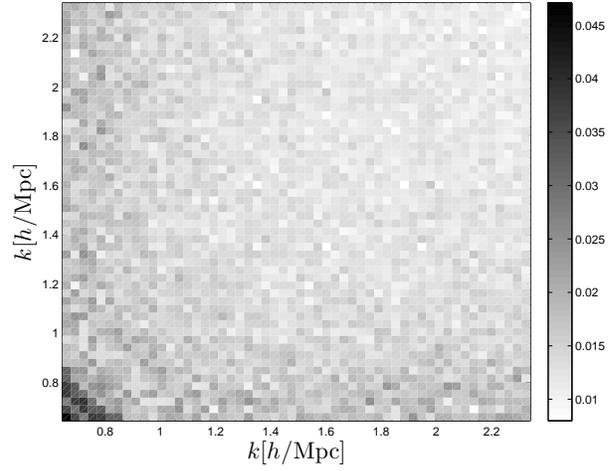}
  \caption{ Fractional error between the $C_{0}$ matrix obtained 
    from the $C(k, k', \theta)$ for all shell combinations, 
    and that obtained directly from the angle averaged $P(k)$. 
    We do not show the largest scales, which are noisy due to low statistics.
    We have also divided the $C_{0}$ matrix by $(4\pi)$ for the two objects to match exactly.}
    \label{fig:C0_matrix_frac_error}
  \end{center}
\end{figure}

By looking at the fractional error between the $C_{0}$ matrix and the actual covariance matrix
of angle averaged power spectra (Fig.  \ref{fig:C0_matrix_frac_error}), 
we find that our method provides  a very good agreement in the trans- and non-linear regimes, 
down to the few percent level. 
We do not  show the largest scales, in which our method is more noisy, for reasons already explained.
We recall that an extra contribution to $C_{0}(k,k')$, not included here, comes from the non-Gaussian Poisson uncertainty,
as discussed in section \ref{subsec:poisson}, and needs to be included in the final analysis.


\begin{figure}
  \begin{center}
 \centering
\includegraphics[width=3.2in]{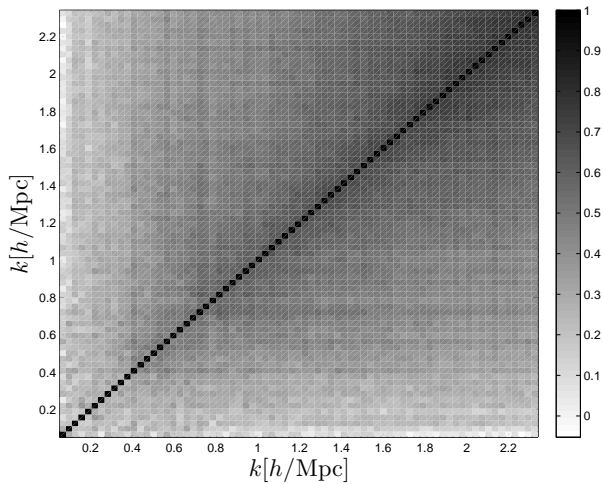}
  \caption{ $C_{2}$ matrix, normalized such that the diagonal elements are equal to unity.
    The off-diagonal elements are still correlated at least at $40\%$ for scales smaller than $k=1.0 h\mbox{Mpc}^{-1}$.}
    \label{fig:C2_matrix}
  \end{center}
\end{figure}

We now  present the next few multipole matrices, and we found that 
beyond $\ell =4$, very little information contained in the off-diagonal elements.
Fig.  \ref{fig:C2_matrix} shows the $C_{2}$ matrix, again normalized to the diagonal
for visual purposes. We observe that the smallest scales are correlated up to $60$ per cent. 
As discussed in sections \ref{subsec:weaklensing} and \ref{sec:discuss}, this matrix is a requirement for an accurate treatment of BAO 
and weak lensing data analyses, 
first because the uncertainties on the underlying galaxy surveys
are not isothropic, and second because the optimal
window function is often also angle-dependent \cite{2010PhRvD..81l3015L}.

\begin{figure}
  \begin{center}
  \centering
\includegraphics[width=3.2in]{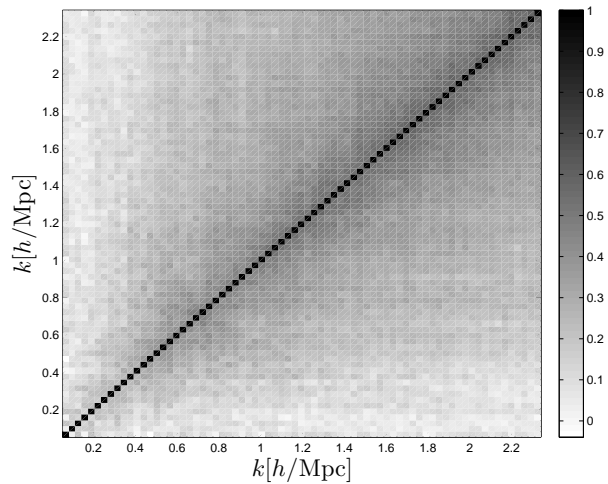}
  \caption{ $C_{4}$ matrix, normalized such that the diagonal elements are equal to unity.
    The off-diagonal elements are correlated at the $30\%$ level in the non-linear regime, and not far off the diagonal.}
    \label{fig:C4_matrix}
  \end{center}
\end{figure}

\begin{figure}
  \begin{center}
  \centering
\includegraphics[width=3.2in]{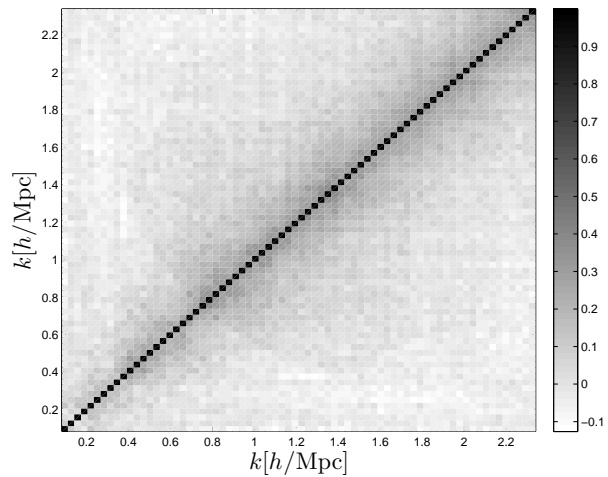}
  \caption{ $C_{6}$ matrix, normalized such that the diagonal elements are equal to unity.
    We observe that the matrix is mostly diagonal, and thus decide to treat $C_{6}$  and all higher multipoles as purely Gaussian. }
    \label{fig:C6_matrix}
  \end{center}
\end{figure}

Fig.  \ref{fig:C4_matrix} shows that the correlation in the $C_{4}$ matrix
is still of the order $50$ per cent for a good portion of the non-linear regime. 
The new feature here is that the strenght of the correlation of strongly non-linear modes 
among themselves starts to decrease as we move away from the diagonal. 
Fig.  \ref{fig:C6_matrix} shows that in $C_{6}$, the matrix is mostly diagonal.
As we progress through higher multipole moments, the off-diagonals become even dimmer,
hence do not contain significant amount of new information. 
From this perspective, a multipole expansion up to $\ell = 4$ is probably as far as one needs to push
in order to medel correctly the off-diagonal elements.

Following [Eq.\ref{eq:cov_expansion}], we thus propose to reconstruct the full $C(k,k',\theta)$ from a
combination of a) fully non-linear $C_{\ell}(k,k')$ matrices (for $\ell \le 4$), presented above, 
b) analytical  terms given by [Eq. \ref{eq:Cl_Gauss}] (which we scale up by $30$ per cent 
 as mentioned in section \ref{subsec:C_ell_den}), and c) non-Gaussian Poisson error, which depends solely
 on the number density od the sampled fields.
In the  next section, we  decompose and simplify these $C_{\ell}$ matrices
into a handfull of fitting functions, and show how one can easilly reconstruct the full $C(k,k',\theta)$
at the percent level precision.

\begin{figure}
  \begin{center}
   \centering
\includegraphics[width=3.2in]{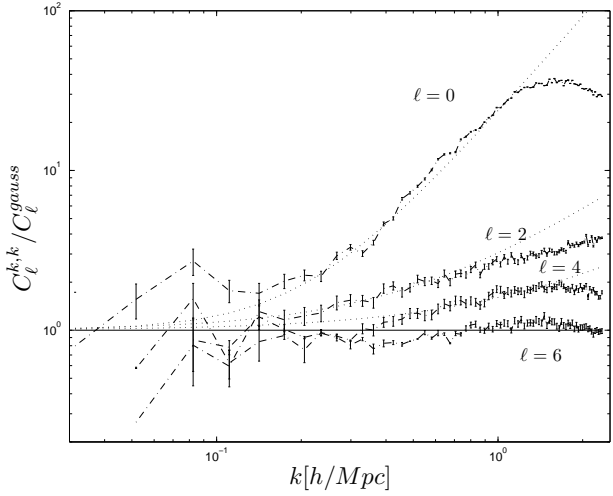}
\caption{Ratio of the diagonal elements of a few $C_{\ell}$ matrices, 
     compared to the Gaussian prediction. The error bars were obtained from bootstrap re-sampling.
     Over-plotted are fitting functions, summarized in Table \ref{tab:fit_ratio}.}
    \label{fig:diag_many_Cl}
  \end{center}
\end{figure}

We next present in Fig.  \ref{fig:diag_many_Cl} the ratio of the diagonal 
of these matrices to the Gaussian prediction.
We observe that all of them are consistent with the Gaussian prediction in the 
linear regime.  As we progress towards the non-linear regime, 
the largest departure comes from the $C_{0}$ matrix, by a factor of about  $40$ near 
$k=1.0 h \mbox{Mpc}^{-1}$. We observe a turn over at smaller scales, which is caused by our resolution limit. 
We opted not  to model it in our fitting formula.
$C_{2}$ and $C_{4}$ mildly break away from Gaussianity by factors of $4$ and $2$ at the same scale.
All the higher $\ell$'s are consistent with Gaussian statistics.
Over-plotted on the figure are fitting formulas, which are summarized in Table \ref{tab:fit_ratio}.

\begin{table}
  \begin{center}
\caption{Fitting formulas for the ratio between the diagonals of the $C_{\ell}(k,k)$ and 
  the Gaussian prediction. For all $\ell$'s, the function is modeled by 
  $V(x) = 1.0 + (x/\alpha)^{\beta}$.}
\begin{tabular}{|c||c|c|c|c|}
\hline \hline
$\ell$  & $\alpha$ & $\beta$ \\
\hline \hline
0&   0.2095  &  1.9980 \\
\hline
2&   0.5481    & 0.7224 \\
\hline
4&   1.6025    & 1.0674 \\
\hline \hline
\end{tabular}
  \label{tab:fit_ratio}
  \end{center}
\end{table}



\section{Factorization of the $C_{\ell}$ Matrices}
\label{sec:factorization} 

In this section, we simplify even further our results
with an Eigenvalue decomposition of the $C_{\ell}(k,k')$ matrices, 
normalized to unity on their diagonal, as shown in the figures of 
section  \ref{subsec:ClMatrix}.
We perform an iterative process to factorize each matrix into
a purely diagonal component and a symmetric, relatively smooth off-diagonal part.
The later can be further decomposed  into a small set of Eigenvectors $U_{\lambda}(k)$,
corresponding to the largest Eigenvalues $\lambda$. 
These are then fitted with simple formulas.
Combined with Gaussian predictions and fitting formulas from the deviation on the diagonal, 
one can fully reconstruct each of the $C_{\ell}(k,k')$ matrix, and thus recover $C(k,k',\theta)$ as well.

\begin{figure}
  \begin{center}
\centering
\includegraphics[width=3.2in]{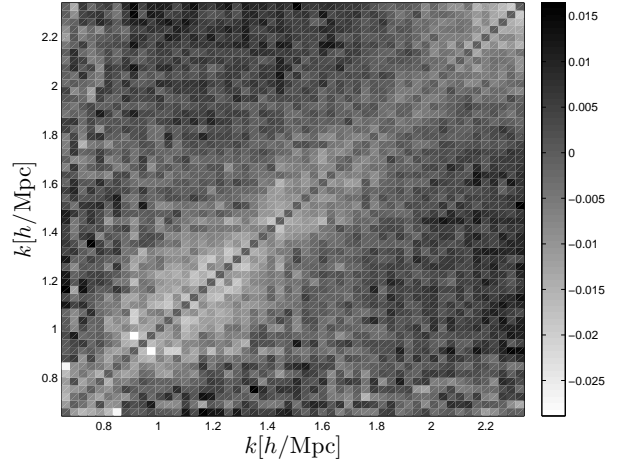}
  \caption{ Fractional error between the original $C_{0}$ matrix and that produced with
    the principal Eigenvector. 
    We do not plot the largest scales, which are noisy due to low statistics.}
    \label{fig:fracError_Eigen_C0}
  \end{center}
\end{figure}

We start off the iteration by assigning 
the identity matrix to the diagonal component, which we subtract from the original
matrix. We then extract from the remainder the principal Eigenvectors 
and recompose an new matrix as 
\begin{eqnarray}
r_{\ell}(k,k') \equiv  \frac{C_{\ell}(k,k')}{\sqrt{C_{\ell}(k,k) C_{\ell}(k',k')}} = \delta_{kk'} + \sum_{\lambda} \lambda U_{\lambda}(k) U_{\lambda}(k')
\label{eq:eigen_decomp_init}
\end{eqnarray}
For the next iteration, we model the diagonal as $\delta_{k,k'} - \sum_{\lambda}\lambda (U_{\lambda}(k))^{2}$,
and decompose the remainder once again. 
We iterate until the results converge, which takes about 4 steps.
We vary the number of Eigenvalues we keep in our iteration, and keep the minimal number 
for which the reconstruction converges.
In the end, the $r_{\ell}(k,k')$ matrix  is modeled as:
\begin{eqnarray}
r_{\ell}(k,k') = \delta_{kk'}\big[1  - \lambda U_{\lambda}^2(k)\big] + \sum_{\lambda} \lambda U_{\lambda}(k) U_{\lambda}(k')
\label{eq:eigen_decomp}
\end{eqnarray}
We show in Fig.  \ref{fig:fracError_Eigen_C0} the fractional error between the original
matrix and the factorized one. The factorization of the $C_{0}$  matrix with one Eigenvector
reproduces the original matrix at the few percent level.
The same procedure is also applied for the higher multipoles, in which we have included the first four Eigenmodes,  
and we find that the fractional error between the reconstructed and the original matrix are also of the order of a few percent.

\begin{figure}
  \begin{center}
    \centering
\includegraphics[width=3.2in]{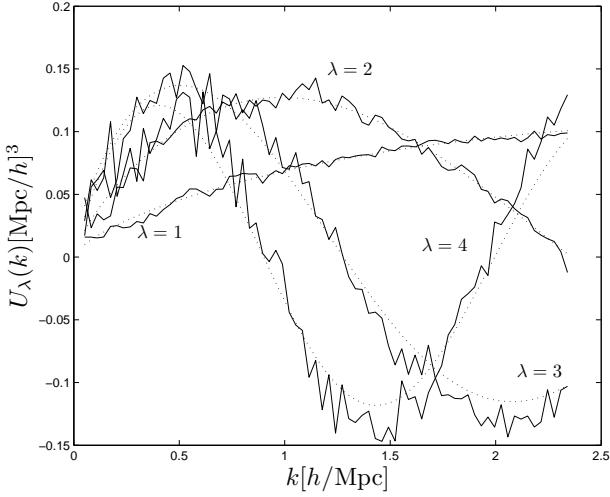}
  \caption{ Four principal Eigenvectors of the $C_{4}$ normalized matrix (solid lines), 
and corresponding fitting formulas (dotted lines).}
    \label{fig:EigenFitC4}
  \end{center}
\end{figure}

We next fit these Eigenvectors with simple functions:
for all $\ell$'s, the first Eigenvector is parametrized as $U(k) = \alpha k^{\beta}(\gamma - \delta k)$,
and all the other vectors as $U(k) = \alpha k^{\beta}\mbox{sin}(\gamma k^{\delta})$.
The values of $(\alpha, \beta, \gamma, \delta) $ for the lowest three $\ell$'s are presented in Table \ref{tab:fitfunction}.
We required that all of these formulas vanish as $k \rightarrow 0$, since the $C_{\ell}$  matrices become diagonal
in the linear regime. 
The Eigenvectors of the $C_{4}$ matrix are presented in Fig.  \ref{fig:EigenFitC4}; over-plotted are the fitting formulas.
The pixel-by-pixel agreement between the original matrices and those obtained from the fitted formulas is with less than 10  percent for all $\ell \le 4$ and $k>0.5$. 

Larger scales fluctuate much more as they are less accurately measured, hence
the pixel-by-pixel agreement is not expected there. In addition, the matrices with $\ell \ge 6$ are much harder
to express with a small set of Eigenvectors, since the Eigenvalues are not decreasing fast enough.
In any case, the first three harmonics we provide here contain most likely all the information one will ever use
in realistic surveys and forecasts.  
   

\begin{figure}
  \begin{center}
\centering
\includegraphics[width=3.2in]{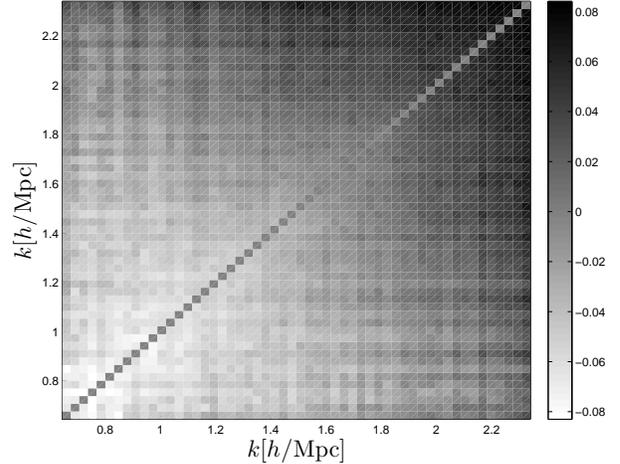}
  \caption{ Fractional error between the original $C_{0}$ matrix and that produced
    with the fitting formulas. 
    We do not show the largest scales, which are noisy due to low statistics.}
    \label{fig:fracError_Fit_C0}
  \end{center}
\end{figure}

\begin{table}
  \begin{center}
\caption{Fitting parameters for the Eigenvectors of the $C_{\ell}$, with their corresponding Eigenvalues.
For all $\ell$'s, the first Eigenvector is parametrized as $U(k) = \alpha  \left(\frac{\beta}{k} + \gamma\right)^{-\delta}$, 
and all the other vectors as $U(k) = \alpha k^{\beta} \mbox{sin}(\gamma k^{\delta})$.
These were obtained from dark matter N-body simulations, but the method is general, and a different 
prescription of galaxy population may result in slightly different values. }
\begin{tabular}{|c||c|c|c|c|c|}
\hline \hline
$\ell$ &$\lambda$ & $\alpha$ & $\beta$ & $\gamma$  & $\delta$\\
\hline \hline
0&   61.9058  & 0.0501 & 0.0207  &  0.6614 &  2.3045 \\
\hline
2&    35.7400 & 0.273 & 0.8266   & 1.962& 0.816     \\
 &     4.4144   & 0.15772  & 2.4207    & 0.79153 &  0.032207 \\
 &      1.7198  & 0.14426   & 4.0613   & 0.76611 & -0.26272     \\
 &      0.9997  & 0.14414  &  5.422  &    0.84826 & 0.31324   \\
\hline
4&  22.0881   & 0.060399  & 0.10344   & 0.64008 & 2.2584    \\
 &     4.5984   & 0.1553  &  2.3370 &  0.9307&  -0.1154   \\
 &     2.2025   & 0.1569  &   3.6937  & 0.92304 &  0.04006  \\
 &     1.4062   & 0.15233  &  5.1617  &  0.8899& -0.14503   \\
\hline \hline
\end{tabular}
  \label{tab:fitfunction}
  \end{center}
\end{table}


\subsection{Non-Gaussian Poisson noise}
\label{subsec:non_gaussian_poisson}

The non-Gaussian Poisson uncertainty, whose construction was presented in section \ref{subsec:poisson},
can conveniently be incorporated in an analysis by finding the corresponding  Eigenvalue and Eigenvectors
of $C_{0}^{Poisson}(k,k')$. Higher multipoles are not relevant as the angular distribution is flat, as shown in the 
middle plot of Fig.  \ref{fig:test_cov}. We tested three number densities, corresponding to 
 $n = 5.0 \times 10^{-5}, 1.52 \times  10^{-4}$ and $1.0  \times  10^{-2} h^{3}/\mbox{Mpc}^{3}$. 
In all cases, we computed the covariance matrix, decomposed 
into a diagonal component and a cross-correlation coefficient matrix, 
found the matrix's leading eigenvalue and Eigenvector,  then fitted the latter with the same fitting function:
$U^{Poisson}_{fit}(k)  = \alpha  \left(\frac{\beta}{k} + \gamma\right)^{-\delta}$.
The diagonal was also fitted with a simple power law of the form 
\begin{eqnarray}
V^{Poisson}(k) \equiv \frac{C^{Poisson}(k,k)}{C^{Poisson}_{Gauss}(k,k)} = e^{\epsilon}k^{\sigma}
\end{eqnarray}
where $C^{Poisson}_{Gauss}(k,k) \equiv \frac{P_{Poisson}^{2}(k)}{N(k)}$.
 The best-fitting parameters are summarized in Table \ref{tab:fitfunction_poisson},
 and the performance of the Eigenvector fit can be found in the Appendix.

 \begin{table}
  \begin{center}
\caption{Fitting parameters for the diagonal of the  $C_{0}^{Poisson}(k,k')$ matrix, and for the
fit of the Eigenvectors of the corresponding cross-correlation coefficient matrix.
For all three number densities (i.e. $n_{1,2,3} =  5.0 \times 10^{-5}, 1.52 \times 10^{-4}$ and $1.0 \times 10^{-2}$  respectively), the Eigenvector is parametrized as 
$U_{fit}^{Poisson}(k) = \alpha  \left(\frac{\beta}{k} + \gamma\right)^{-\delta}$, 
and the ratio of the diagonal to the Gaussian prediction is fitted with
$V^{Poisson}(k) = e^{\epsilon}k^{\sigma}$. Top to bottom corresponds to increasing density.}
\begin{tabular}{|c|c|c|c|c|c|c|}
\hline \hline
 $\lambda$ & $\alpha$ & $\beta$ & $\gamma$  & $\delta$ & $\epsilon$ & $\sigma$ \\
\hline \hline
   52.02 & 1.0193 & 0.0947 & 2.1021 & 2.5861 & 2.6936 & 2.1347\\
\hline
     45.09 & 0.9987 & 0.2034 & 2.1553 & 2.3407 & 1.6533 & 2.1965  \\
\hline
  24.41 & 0.2966 & 3.3736 & 0.6099 & 0.6255 & -0.4321 & 2.0347 \\
\hline \hline
\end{tabular}
  \label{tab:fitfunction_poisson}
  \end{center}
\end{table}

\subsection{Recipe}

Here we summarize our method to generate accurate non-Gaussian matrices.
The full $C(k,k',\theta)$ matrix is then given by [Eq. \ref{eq:cov_expansion}],
where the $\ell \le 4$ terms are obtained from the fitting functions, and the 
higher multipole moments are obtained directly from [Eq. \ref{eq:Cl_Gauss}].
The sum over these Gaussian terms can be evaluated analytically as 

\begin{eqnarray}
    \frac{1}{2}\sum_{\ell  =6}^{\infty}  (2\ell + 1)(1 + (-1)^{\ell}) P_{\ell}(\mu) &=& \lefteqn{\delta_{D}(1+\mu)  + \delta_{D}(1-\mu) -  }\nonumber \\
     &&1 - 5 P_{2}(\mu) - 9 P_{4}(\mu) 
\end{eqnarray}

For the non-Gaussian terms, we proceed as follow:
Each of the $C_{\ell}(k,k')$ matrices, normalized on the diagonal, can be constructed from
the first set of fit functions $U_{\lambda}(k)$ provided in Table \ref{tab:fitfunction},
by  following  [Eq. \ref{eq:eigen_decomp}].
The `un-normalized' $C_{\ell}(k,k')$ terms are then constructed by inverting 
[Eq. \ref{eq:crosscorr}], where the diagonal elements are obtained from the product  of
the $V_{\ell}(k)$, also summarized in  Table \ref{tab:fitfunction}, with  the Gaussian prediction [Eq. \ref{eq:Cl_Gauss}].
In other words: 
\begin{eqnarray}
   C_{\ell}(k,k')  &=&  \lefteqn{\bigg(\delta_{kk'}\bigg(1- \sum_{\lambda}\lambda  U_{\lambda,\ell}^{2}(k)\bigg)  + \sum_{\lambda}\lambda U_{\lambda,\ell}(k)U_{\lambda,\ell}(k')\bigg) \times}   \nonumber\\
    &&\sqrt{V_{\ell}(k)V_{\ell}(k')C_{\ell}^{Gauss}(k)C_{\ell}^{Gauss}(k')} 
\end{eqnarray}
The complete covariance matrix is given by:
\begin{eqnarray}
   \lefteqn{C(k,k',\mu) = \frac{1}{4\pi} \sum_{\ell  =0}^{3}  (2\ell + 1) C_{\ell}(k,k') P_{\ell}(\mu)   +}  \nonumber\\
   && \frac{2 P(k)^{2}}{N(k)}\bigg(   \delta_{D}(1+\mu)  + \delta_{D}(1-\mu) - 1 - 5 P_{2}(\mu) - 9 P_{4}(\mu)  \bigg)  
\label{eq:full_cov_model}
\end{eqnarray}
with $\mu = \mbox{cos}(\theta)$.
This can be written in a more compact form as
\begin{eqnarray}
C(k,k',\mu) &=& C_{Gauss}(k)\delta({\bf k} - {\bf k'}) +  \nonumber \\
 && \sum_{\ell  =0}^{3}  (2\ell + 1)\bigg(G_{\ell}(k)\delta(k-k') +H_{\ell}(k,k') P_{\ell}(\mu)    \bigg) 
\label{eq:full_cov_model2}
\end{eqnarray}
with
\begin{eqnarray}
G_{\ell}(k) = C_{Gauss}(k)(V_{\ell}(k)- 1)
\end{eqnarray}
\begin{eqnarray}
H_{\ell}(k) = \sum_{\lambda} \bigg(F_{\lambda,\ell}(k) F_{\lambda,\ell}(k') - F_{\lambda,\ell}^{2}(k)\delta(k-k')\bigg)
\end{eqnarray}
and
\begin{eqnarray}
F_{\lambda,\ell}(k) = U_{\lambda,\ell}(k)\sqrt{\lambda V_{\ell}(k) C_{Gauss}(k)}
\end{eqnarray}

We conclude this section with a word of caution when using the fitting formulas provided here,
in the sense that the range of validity of the fit has not been tested on other cosmological volumes.
Consequently, we advice that one should limits itself to $k\le 2.0 h\mbox{Mpc}^{-1}$.

\section{Measuring the Impact with Selection Functions}
\label{sec:surveys}

This section serves as a toy model for a realistic non-Gaussian error analysis,
as it incorporates the non-Gaussian covariance matrix measured from N-body simulations 
with the 2dFGRS selection function  \citep{2002MNRAS.336..907N}. 
We compare the estimated error bars on $P(k)$ between the naive, purely diagonal, Gaussian covariance matrix,
the effect of the one dimensional window function as prescribed by the FKP formalism, 
the unconvolved non-Gaussian covariance as measured from our $200$ N-body simulations, 
and the convolved non-Gaussian matrix.

We recall that in a periodic volume, a selection function that is exactly uniform throughout the volumes makes the 
observed and true covariance matrices exactly equal.
That is only true in simulated volumes, and in that case, the optimal estimator for the power spectrum covariance matrix
is obtained from the non-Gaussian covariance matrix directly measured from the simulations.

Non-periodicity is best dealt with by zero-padding the observed survey, which results in
some coupling between different measure power spectrum bands.
The coupling becomes more important as the selection function departs from a top hat,
and in that case, the best estimator of the observed covariance matrix is 
a convolution of the 6-dimensional covariance  over both vectors $({\bf k}, {\bf k'})$, given by: 
\begin{eqnarray}
C_{obs}({\bf k}, {\bf k'})  = \frac{\sum_{{\bf k''}, {\bf k'''}} C_{true}({\bf k''}, {\bf k'''}) 
|W({\bf k} - {\bf k}'')|^{2} |W({\bf k'} - {\bf k}''')|^{2}}{(N^{2}N_{c}\sum_{{\bf x}}W^{2}({\bf x})w^{2}({\bf x}))^{2}}
\label{eq:non_gauss_cov_est2}
\end{eqnarray}
The denominator is straight forward to calculate, while the numerator is 
a 6-dimensional integral, which must be calculated at all of the 6-dimensional coordinates,  
a task is computationally impossible to perform.
For example, assuming that the size of the grid is $n^3$ cells,
then for each $({\bf k},{\bf k'})$ pair, we have to sum over $n^{6}$ terms.
We also have $n^{6}$ such pairs, and each term takes about $3$ {\tt flop}.
For $n=100$, this process would take  $3 * 10^{24}$ {\tt flop},
and current supercomputers are in the regime of resolving $10^{12}$ {\tt flop} per seconds.
The  above calculation would therefore take about 3000 years to complete. 
With the factorization proposed in this work however, it is now possible to break down the computation into smaller 
pieces  and reduce the loops to 7 dimensions at most.

\subsection{Factorization of the 6-dimensional covariance matrix}

We propose here a break down of the true covariance matrix $C({\bf k}'',{\bf k}''')$
into a product of simple functions of the form $H_{\ell}(k'')$, $G_{\ell}(k'')$ and $P_{\ell}(\mu)$
where the angular components come exclusively from the Legendre polynomials.
As explained in section \ref{subsec:non_gauss_error},  the argument of the Legendre polynomials, $\mu$, 
is the (cosine of ) angle between ${\bf k}''$ and ${\bf k}'''$, which must first be expressed in terms of $(\theta'', \phi'', \theta''',  \phi''')$,
following [Eq. \ref{eq:cos_gamma}]\footnote{In this section, we 
use $\mu$ instead of $\mbox{cos}(\theta)$ to denoted the (cosine of the) angle between the two Fourier modes, to avoid confusion with $\theta'$ and $\theta'''$, 
which corresponds to the angle of $({\bf k}'',{\bf k}''' )$ with respect to the $x$-axis.}.
The only  multipoles that appear in our equations are $\ell = 0,2,4$, so $\mu$ is to be expanded at most up to 
the fourth power. For a full factorization, the terms including $\mbox{cos}(\phi'' - \phi''')$ must further be re-casted in their exponential form with Euler's formula.

When computing the convolution, the first term on the right hand side of [Eq.\ref{eq:full_cov_model2}]
is spherically symmetric, hence it must be convolved with the selection function as:
\begin{eqnarray}
C_{Gauss}^{obs}(k,k') = \sum_{k''}C_{Gauss}(k'')  |W(k''-k)|^{2}|W(k'' - k')|^{2}
\label{eq:Obs_Cov_Gauss}
\end{eqnarray}
which is pretty much the FKP prescription, namely that the selection function
is the only cause of mode coupling.

When computing the other terms  of [Eq.\ref{eq:full_cov_model2}], which are the non-Gaussian contributions, 
we use the fact that the only coupling between the ${\bf k}''$ and ${\bf k}'''$ vectors comes from the delta function,
which couples their radial components.
This means that all the angular integrations can be precomputed and stored in memory. 
For example,  the only angular dependence in the $\ell=0$ multipole comes from the selection function itself, 
hence we can precompute 
\begin{eqnarray}
    X({\bf k}, k'') = \sum_{\theta'',\phi''} |W({\bf k} - {\bf k''})|^{2} \mbox{sin}(\theta'') w(\theta'',\phi'')
 \label{eq:X}
\end{eqnarray}
and the convolution is now four dimension smaller. 
The weight function $w(\theta'',\phi'')$ is equal to unity for the $C_{0}$ term, 
and the $\mbox{sin}(\theta'')$ comes in from the Jacobian in  angular integration.
For the higher  multipoles, more of such terms must be precomputed as well, whose weight functions are summarized in Table \ref{table:Xterms}.

\begin{table}
  \begin{center}
  \caption{List of weights $w(\theta,\phi)$ needed for the angular integrals over the selection function. These can be precomputed to speed up the convolution.
All integrals are in the form of [Eq.\ref{eq:X}].}
  
\begin{tabular}{|c|c|c|c|}
\hline 
$\mbox{cos}^{2}(\theta)$ & $\mbox{sin}^{2}(\theta)$ & $\mbox{cos}^{2}(\theta) e^{\pm 2 i \phi}$ & $\mbox{sin}^{2}(2 \theta)e^{\pm i \phi}$  \\
\hline
$\mbox{cos}^{4}(\theta)$ & $\mbox{sin}^{4}(\theta)$ & $\mbox{sin}^{4}(\theta) e^{\pm 2 i \phi}$ & $\mbox{sin}^{4}(2 \theta)e^{\pm 4 i \phi}$  \\
\hline
$\mbox{sin}^{2}(2 \theta)$ & $\mbox{sin}^{2}(2 \theta) e^{\pm 2 i \phi}$ & $\mbox{sin} (\theta) \mbox{cos}^{3}  (\theta) e^{\pm i \phi}$ & $\mbox{cos} (\theta) \mbox{sin}^{3}  (\theta) e^{\pm i \phi}$ \\
\hline
\end{tabular}

\label{table:Xterms}
\end{center}
\end{table}


\subsection{The 2dFGRS selection function}

The 2dFGRS\citep{2003astro.ph..6581C} is comprised of two major regions, the NGP and the SGP,
each of which take the overall form of a fan of $75 \times 5$ degrees, extending from $z=0.02$ to $ 0.22$.
The selection function  is constructed by first integrating the luminosity function $d\Phi(L)/dL$ 
over all the observed luminosity range, which is both redshift and 
angle dependent, and multiplying the result by the redshift completion function $R(\theta,\phi)$. Namely, we define the galaxy number density $\tilde{n}(z,\theta,\phi)$ as:
\begin{eqnarray}
\tilde{n}(z,\theta,\phi) = \int_{y_{min}(z,\theta,\phi)}^{\infty}  \frac{d\Phi(y)}{dL} y dy
\label{eq:gal_den}
\end{eqnarray}
where $y=L/L^{\star}$,  $L^{\star}$ being the characteristic galaxy luminosity, and  $\frac{d\Phi(y)}{dL}$ given by
\begin{eqnarray}
\frac{d\Phi(y)}{dL} = \Phi^{\star}y^{\alpha + 1}e^{y}
\label{eq:phi}
\end{eqnarray}
The three parameters $\Phi^{\star}$, $\alpha$ and $M_{\star} - 5log_{10}h$ are obtained from the 2dFGRS as $-1.21$, $1.61x10^{-2} h^{3}\mbox{Mpc}^{-3}$ and $-19.66$ respectively.
The integral over $y$ gives an incomplete gamma function:
\begin{eqnarray}
\tilde{n}(z,\theta,\phi) = \Phi^{\star}\Gamma(\alpha+2, y_{min}(z,\theta,\phi))
\label{eq:gal_den_2}
\end{eqnarray}
in which the term $y_{min}$ can be expressed as 
\begin{eqnarray}
\mbox{log}_{10}(y_{min}(z,\theta,\phi) ) &=&\lefteqn{ 0.4 \bigg( M_{\star} - 5 \mbox{log}_{10}h   - b_{J}(\theta,\phi)  +}   \nonumber\\
&& 5 \mbox{log}_{10}\left(\frac{D_{L}(z)}{10\mbox{pc}/h}\right) + \frac{z+ 6z^{2}}{1 + 20z^{3}}\bigg)
\label{eq:y_min}
\end{eqnarray}
where $b_{J}(\theta,\phi)$ is the angular  dependence of the magnitude sensitivity, $D_{L}(z)$ is the luminosity distance that is used to convert between
absolute relative magnitudes, and the last term in the right hand side is the $K-$and $e-$corrections.  
Finally, the selection function is simply :
\begin{eqnarray}
W(z,\theta,\phi) = \tilde{n}(z,\theta,\phi)R(\theta,\phi)
\label{eq:W}
\end{eqnarray}
Both $R(\theta,\phi)$ and $b_{J}(\theta,\phi)$ are publicly available from the 2dFGRS website\footnote{www.mso.anu.edu.au/2dFGRS/}.
It is possible to obtain an even more accurate selection function by taking into account the redshift dependence of the 
magnitude  sensitivity, however we  do not need such an accuracy for the current work. 
We finally normalize the selection function such that 
\begin{eqnarray}
\int |W({\bf k})|^{2} d^3k = 1
\label{eq:nrm_wk}
\end{eqnarray}

To understand the impact of the non-Gaussian Poisson uncertainty on the measured uncertainty,
we test various templates, keeping the 2dFGRS selection function fixed. We follow the procedure of section \ref{subsec:poisson}, 
with an average  number density of $n_{gal}=1.52 \times 10^{-4} h^{3}\mbox{Mpc}^{-3}$,
which corresponds to an early data release of the 2dFGRS data. The final release contains more objects, and has a density
of about  $n=5.0 \times 10^{-2} h^{3}\mbox{Mpc}^{-3}$.
By comparison, the Poisson uncertainty corresponding to the number count of the Wiggle-Z survey could be modeled with 
$n=5.0 \times 10^{-5} h^{3}\mbox{Mpc}^{-3}$ for partial data and about $2.0 \times 10^{-4} h^{3}\mbox{Mpc}^{-3}$ for the final data release.
We thus opt for two more number densities:  $n  =  1.52 \times 10^{-4}$ and $n  =  1.0 \times 10^{-2}$.

\subsection{Results}

\begin{figure}
  \begin{center}
\centering
\includegraphics[width=3.2in]{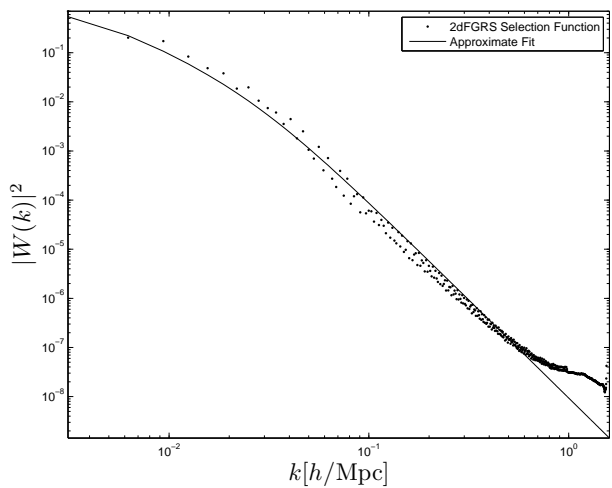}
  \caption{The angle average of the 2dFGRS selection function, compared to an approximate fit provided by \citep{2001MNRAS.327.1297P}.
  The fit is not perfect as it was obtained with an earlier estimate of the selection function. 
  We also note that our method differs in details with that used in \citep{2005MNRAS.362..505C} by that fact that we imposed a cut 
  at redshift of $z=0.22$, and that we used a somewhat smaller resolution. }
    \label{fig:Wk}
  \end{center}
\end{figure}

We assign the selection function on to a 256x256x128 grid, where the lower resolution is along the direction perpendicular from the NGP.
We precompute the Fourier transform, $W({\bf k})$ and square each terms. 
Fig.  \ref{fig:Wk} shows  a comparison between the angle average of $|W({\bf k})|^{2}$  and a fitting function
provided by the 2dFGRS group.

We then define a second set of bins in spherical coordinates, over which we perform the convolution. 
For that purpose, we divide the original volume of the survey into 64 radial bins, 48 polar bins and 32 azimuthal bins.
The selection function is assigned in the grid by averaging over the 27 closest cells in the original grid.
We have included a $\mbox{sin}(\theta)$ terms in each integrals over the polar angle, and a $k^2$ in each 
radial integral to properly account for the Jacobian matrix in spherical coordinates.

\begin{figure}
  \begin{center}
   \centering
\includegraphics[width=3.2in]{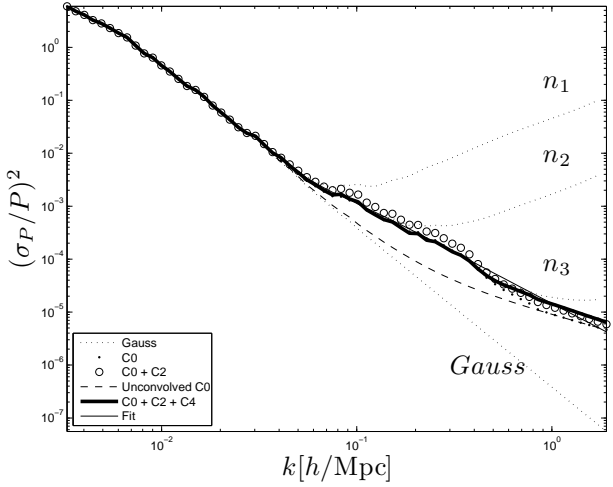}
  \caption{ Diagonal of the convoluted covariance matrix, first with no multipole i.e. following FKP prescription (bottom-most dotted line), 
  then with the progressive inclusion of the $C_{0}$ (solid black points),
  the $C_{2}$ (open circles) and the $C_{4}$ multipoles (thick solid line). Also shown is the diagonal of the unconvolved $C_{0}$ terms
  directly measured from N-body simulations (dashed line), and a fitting function for the total covariance (thin solid line). 
  Finally,  the inclusion of the non-Gaussian Poisson noise is represented by three dotted lines,
  representing the three number density detailed in Table \ref{tab:fitfunction_poisson}.
  The 2dFGRS final data release has a number density of the order  $5.0 \times 10^{-2} h^{3}\mbox{Mpc}^{-3}$, 
  which thus lies between  $n_{2}$  and $n_{3}$.}
    \label{fig:ConvolvedDiag}
  \end{center}
\end{figure}

Fig.  \ref{fig:ConvolvedDiag} shows the diagonal of the convolved covariance matrix, divided by $P^{2}(k)$, for the FKP prescription
and for the progressive inclusion of $\ell=0,2$ and $4$ multipoles. Also overploted is the 
non-Gaussian results without the convolution.
We see that already at $k\sim 0.1 h\mbox{Mpc}^{-1}$, the non-Gaussian fractional error, after the convolution,
deviates from the FKP prescription by a factor of about 3.0, while the unconvolved $C_{0}$ 
still traces quite well the FKP curve. This means that the mode mixing caused by the convolution
with the survey selection function increases significantly the variance of the observed power spectrum.
The departure gets amplified as one progresses towards higher $k-$modes, and by $k \sim 1.0$,
the unconvolved $C_{0}$ departs from the FKP prescription by almost two orders of magnitudes.
Interestingly, the convolved $C_{0}$ merges with the unconvolved counterpart at $k\sim0.5$, 
where the BAO scale is usually cut off. inclusion of higher multipole increases the variance by 
a factor of about 2.0. 

We have overplotted a simple smooth fitting function of the form :
\begin{eqnarray} 
C_{fit}(k) =  C_{g}(k)(1 + \frac{2.3}{(0.08/k)^{3.7} + (0.08/k)^{1.1}} + 0.0007)
\label{eq:fit_diag_C4}
\end{eqnarray}
which approximates the contribution from the three lower multipoles.

\begin{figure}
  \begin{center}
    \centering
\includegraphics[width=3.2in]{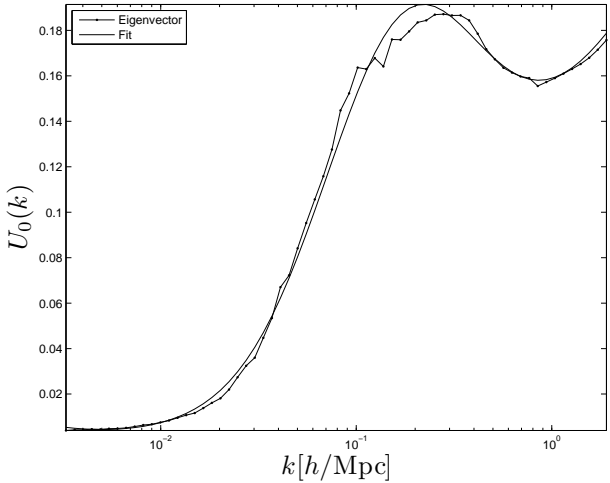}
  \caption{ Principal Eigenvector of the convolved $C_{0}$ matrix, compared to a simple fitting formula.
  The fractional error of the fitting function is at most $13$ per cent. }
    \label{fig:EigenU_obs}
  \end{center}
\end{figure}

\begin{figure}
  \begin{center}
\centering
\includegraphics[width=3.2in]{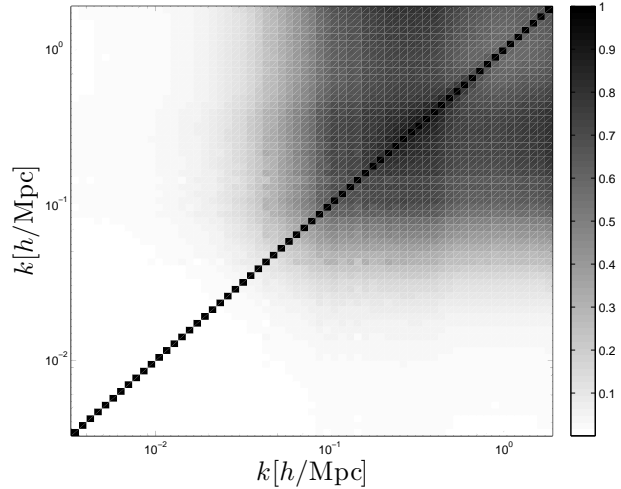}  
\caption{ Normalized convoluted covariance matrix with all three multipole.}
    \label{fig:ConvolvedCovMatrix}
  \end{center}
\end{figure}

\begin{figure}
  \begin{center}
     \centering
\includegraphics[width=3.2in]{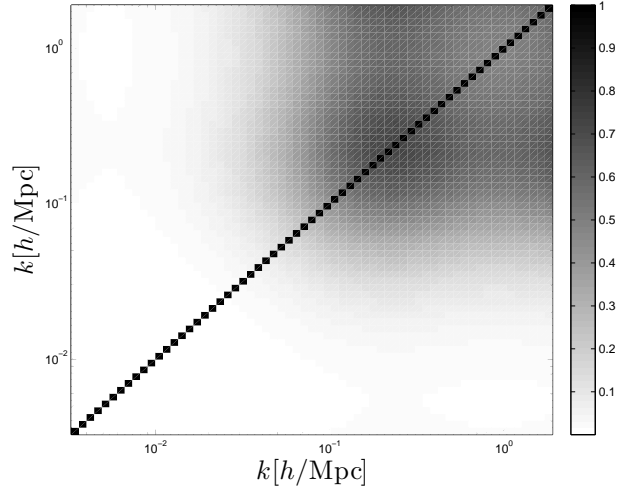}
  \caption{ Normalized convoluted covariance matrix with all three multipole, 
  reconstructed from a fit of the principal Eigenvector. }
    \label{fig:ConvolvedCovMatrixEigen}
  \end{center}
\end{figure}

Fig.  \ref{fig:ConvolvedCovMatrix} shows the convolved cross-correlation coefficient matrix,
where the angle average has been taken after the convolution. 
It is also possible to factorize this matrix, hence we proceed to an Eigenvalue decomposition,
following the same iterative procedure as  in section \ref{sec:factorization}, 
solving for the first Eigenvector only. The Eigenvalue was found to be $\lambda = 19.7833$,
and we used the sum of a quadratic and a Gaussian function  to model the Eigenvector:
\begin{eqnarray}
U_{\lambda}^{obs}(k) &=&\lefteqn{ A\mbox{exp}[-\frac{1}{\sigma^2}\mbox{log}^{2}\left(k/k_p\right)]  +  }\nonumber\\
& &               (a\mbox{log}^{2}\left(k/k_o\right) + b\mbox{log}\left(k/k_o\right) + c) \nonumber \\             
\end{eqnarray}
with $A =  0.1233, \sigma = 1.299, a = 0.0049, b = 0.0042, c = 0.0052$ and $(k_p,k_o) = (0.17,0.008) h\mbox{Mpc}^{-1}$ respectively.
A comparison of the fit and the actual vector is presented in Fig.  \ref{fig:EigenU_obs}.
The noise reduced cross-correlation coefficient matrix is presented in Fig.  \ref{fig:ConvolvedCovMatrixEigen}.
We observe that the Fourier modes are already more than $50$ per  cent correlated at $k = 0.1  h\mbox{Mpc}^{-1}$,
an significant enhancement compared to the unconvolved $C_{0}$ matrix, 
in which the equivalent coupling occurs roughly towards $k = 0.22  h\mbox{Mpc}^{-1}$.
This would most likely have an impact on a non-Gaussian BAO analysis.

\subsection{Applications to  weak lensing}
\label{subsec:weaklensing}

The results presented in  section \ref{subsec:ClMatrix} and the recipe
presented in the previous section can find useful applications
in the field of weak lensing. 
Convergence maps, for instance, are constructed from a redshift integral over
a past line cone filled with dark matter, weighted by a geometric kernel.
Because of the projection nature of this process, the survey maps are sensitive
to both large and small scales, where non-Gaussianities have been observed in the convergence power spectrum \citep{2009arXiv0905.0501D}.

It has recently been demonstrated that weak lensing of  high redshift sources by  large scale structures
can serve as a probe of the dark energy equation of state \citep{2002PhRvD..65f3001H}. 
To make a complex story short, the weak lensing  power spectrum is closely related to that of the three dimensional dark matter 
via Limber's approximation \citep{1954ApJ...119..655L}, hence it is similarly  sensitive to cosmological parameters.
It was then realized that the same structures were also distorting the signal of high redshift source 
of 21 cm emission \citep{2009MNRAS.394..704M},
hence the detection of weak lensing from diffuse temperature fields
could also constrain the dark energy.
The lensing fields are quadratic functions of smoothed temperature fields,  
and the optimal smoothing window function depends not only on the the parameter under
study, but also on the statistical nature of the source and lenses \citep{2008MNRAS.388.1819L}.
 
Optimal quadratic estimators of lensing fields
were first obtained under the Gaussian assumption \citep{2002ApJ...574..566H,2006ApJ...653..922Z},
but it was soon realized that when the observed field is non-Gaussian, the proposed estimators are no longer  optimal:
the noise estimation can be underestimated by several orders of magnitude \citep{2008MNRAS.388.1819L}.
Optimal  non-Gaussian lensing  estimators were then obtained from N-body simulations \citep{2010PhRvD..81l3015L}, and it
was found that the optimal smoothing window function for dark energy involves 
the first two multipoles of the dark matter power spectrum covariance matrix, $C_{0}(k,k')$ and $C_{2}(k,k')$ 
(see [Eqs. $23-24$] in  \citep{2010PhRvD..81l3015L}).
The tools developed in the present paper thus allow one to construct, for the first time and from simple fitting functions, 
optimal non-Gaussian estimators of dark energy parameters from 21 cm temperature maps.

The survey selection function was factored out of the above  discussion,
which was interested solely in the non-Gaussianities inherent in the sources and lenses.
However, cosmology from weak lensing observations of galaxy and quasar surveys is affected by selection functions,
in a similar way to BAO analyses discussed in this paper:
the underlying 2- and 4-point functions in Fourier space are inevitably convolved with the survey selection function.
With the factorization proposed in the paper, this convolution should also be relatively simple to 
perform, and allows one to measure non-Gaussian error bars on the lensing power spectrum,
which includes the effect of the non-linear dynamics, of the selection function, and possibly of
Poisson noise.

In particular, the weak lensing convergence field ${ \bf \kappa}$ is obtained from the dark matter field ${\bf \delta}$ from
\begin{eqnarray}
{\bf \kappa} = \int_{0}^{\chi_{s}} w(\chi, \chi_{s}) {\bf \delta} d\chi
\label{eq:kappa}
\end{eqnarray}
with
\begin{eqnarray}
w(\chi, \chi_{s})=  \frac{3 H_{0}^{2} \Omega_{m}}{2 c^{2}} \chi \left(1 - \frac{\chi}{\chi_{s}} \right) (1 + z(\chi))
\label{eq:kernel}
\end{eqnarray}
$H_0$ is the Hubble parameter, $c$ the speed of light,  $\chi$ and $\chi_s$ the comoving
distances to the lenses and to the source respectively.
When one has knowledge of the three dimensional selection function, it 
can be incorporated in the above equation as
\begin{eqnarray}
{\bf \kappa(\theta,\phi)} = \int_{0}^{\chi_{s}} w(\chi, \chi_{s}) {\bf \delta(\chi, \theta,\phi)} W(\chi, \theta, \phi) d\chi 
\label{eq:kappa_selection}
\end{eqnarray}
we can be further compactified if one absorbs $w$ in the definition of the selection function.
The result takes the form of an integral over one dimension of an observed density fields:
$\kappa(\theta,\phi) = \int \delta({\bf x}) W({\bf x}) d\chi$. 
This means that all the non-Gaussian calculations performed on the three dimensional density fields (i.e. see section \ref{subsec:non_gauss_error})
can be extended to weak lensing maps by the simple  modification of the selection function mentioned above,
plus an integration over the third dimension (in the small angle approximation at least).
In particular, the best estimate of the non-Gaussian lensing covariance matrix
is obtained from solving [Eq. \ref{eq:non_gauss_cov_est}] with the modified selection function,
then by integrating the two $k_z$-components with, at each integration step, 
the inclusion of an extra weight $\chi$ for the conversion
of $k$-modes to $\ell$-modes.

\section{Discussion}
\label{sec:discuss}

 We have found that even for modes of $k\sim  0.1 h\mbox{Mpc}^{-1}$, the non-Gaussian error bars are higher than those prescribed by the FKP method
 by a factor of a few, due to mode coupling caused by the convolution of the selection function.
 This has to be put in contrast with results from pure N-body simulations, which show that the departure from Gaussianity
 reaches this sort of amplitudes at higher $k$-modes, as seen from Fig.  \ref{fig:ConvolvedDiag}.
 We also observe that with the 2dFGRS, the non-Gaussian Poisson noise plays an important role
 if the number density is smaller than $0.01 h^3 \mbox{Mpc}^{-3}$, but is not enough to characterize all of the non-Gaussian 
 features of the density field. The $C_{0}$ term is the leading contribution of the enhancement observed
 in the range $k = 0.06 -- 0.4 h \mbox{Mpc}^{-1}$, but for larger $k$-modes, $C_{2}$ and $C_{4}$ both
 play an important role.

 Without the convolution, keeping only the $C_{0}$ term, and assuming that  
 the BAO measurement was performed with a non-Gaussian estimator,
 the propagation of the non-Gaussianities on to the BAO dilation scale produces very similar 
 constraints \citep{2011ApJ...726....7T}. 
 The estimators that are used in the data analyses however are Gaussian, while the  
 power spectrum covariance matrices that enter the calculations are either Gaussian  or obtained with mock catalogs. 
 As pointed out previously  \citep{2011arXiv1106.5548N}, the estimators constructed in such a way are 
 inconsistent and should be noise weighted.
 When correcting for that effect, it was found that the consistent -- but suboptimal -- error bars are 
 about $10$ per cent higher than those obtained assuming an optimal estimator. 
 In the light of the current results, the observed (i.e. convolved) covariance matrix is even less Gaussian,
 and it is not obvious that the error on the BAO dilation scale  will be unaffected by 
 this new estimates, since our measurement show significant deviations at scales as large as $0.1 h \mbox{Mpc}^{-1}$. 


It is worth mentioning again that the measurement of the $C_{0}(k,k')$ from $C(k,k',\theta)$ provides an alternative
way to extract the covariance matrix of the angle average power spectra. 
Although the mean value of both methods is identical, i.e. unbiased, the second gives us a better handle
on the error on each matrix element, hence provide an optimal measurement
of their uncertainty. We have shown in this paper that 
each matrix element receives its dominant contribution from small angles, while larger angles
are more noisy. It is thus in possible to re-weight the sum by taking this new information into account, 
and obtain more accurate error bars on each matrix element, compared to the current bootstrap  method (HDP2).
As mentioned in the introduction, our next objective is to achieve a similar accuracy with a much lower number 
of simulations. This would revolutionize the field of observational cosmology as the covariance matrix would
be measured internally,  i.e. directly from the data.

The techniques presented in this paper call for extensions, 
as we did not include redshift distortions nor shot noise in our analysis.
The latter will become important when repeating this procedure on haloes,
and it was shown \citep{Neyrinck:2006xd} that the Fisher information in haloes
is also departing from Gaussianity.
It is straight forward  to perform a similar analyses with a quadratic halo model, 
where the halo density is parametrized by $\delta_{halo}(x) = A \delta(x) + B \delta^{2}(x)$.
This involves an extra cross correlation between the linear and quadratic term, 
and leaves some room for the choice of $A$ and $B$, and ultimately, one should work straight from a halo catalog.
The optimal estimator should also be based on a cosmology independent 
model of the covariance matrix, hence one should compute how the fitting functions scale with $\Omega_m$ and $\omega$. 

As mentioned earlier, the effect of the selection functions is enhanced for survey geometries
that are different from top-hats, and it would be interesting to repeat some of the BAO data analyses that were performed
on such surveys, like the 2dFGRS, Wiggle-Z. 
The current method also applies to surveys with irregular geometries like those obtained from the Lyman-$\alpha$ forest \citep{2007PhRvD..76f3009M,2011MNRAS.415.2257M},
and we are hoping that it will considered in the elaboration of these  future analysis pipelines.

We leave it for future work to match our results with predictions from higher order perturbation theory.
We would like to verify that the angular dependence we observe in the covariance matrix 
is predicted by a complete 4-points function analysis, at least in the trans-linear regime.  
In addition, the extraction of non-Gaussian error bars from two dimensional angular clustering could also be 
performed with techniques similar to those employed here. 

\section{Conclusion}
\label{sec:conclusion}

Estimating  accurately the non-linear covariance matrix of the matter power spectrum  is essential when constraining 
cosmological parameters including, but not restricted to, the dark energy equation of state $\omega$.
So far, many BAO analyses from galaxy surveys were performed under the assumption that the underlying density field 
is Gaussian, which yields a {\it suboptimal}  measurement of the mean power spectrum (and thus of the BAO dilation scale), 
and, at least as important, the error bars are biased. 

To estimate unbiased error bars on the dilation scale is a challenging task but can now be done. 
In the simple case of periodic volume, it was shown recently \citep{2011arXiv1106.5548N} 
that, first, an unbiased error bar on a suboptimal measurement of the mean could be obtained 
from the knowledge of the underlying covariance matrix.
Second, if one did measure optimally the mean BAO dilation scale, 
then the optimal measurement of the error requires an estimate  of the
{\it inverse} of the power spectrum covariance matrix.
This is much more challenging due to the presence  of noise,
even when dealing with simulations embedded in periodic volumes,
but improves the constraining performance by a significant amount. 

When estimating the power spectrum and its uncertainty from data,  the survey selection function
complicates the calculations since the observed quantities are actually convolved
with the selection function. Since the covariance matrix is not isotropic,  
as it depends on  the relative angle between two Fourier modes,
the convolution cannot be simply factored into two radial components.
Hence we are left with a challenging  six-dimensional integral to perform,
which so far has been an unresolved problem.

In this paper, we present a method to perform this convolution  for an arbitrary galaxy survey selection function,
that thus allows one to measure unbiased  error bars on the matter power spectrum.
The estimate is still suboptimal, unless one combines our tools with the PKL  formalism,
but we have nevertheless removed the bias on the error bar.

From an ensemble of 200 N-body simulations, we have measured the angular dependence of 
the covariance of the matter density power spectrum. We have found that on large scales, there is only a weak dependence,
consistent with the Gaussian aspect of the fields in that regime.
On smaller scales, however, we have detected a strong signal coming from Fourier modes separated by 
small angles. This comes from the fact that the complex phases of these modes are similar,
hence  they tend to couple first.
We next  expanded the covariance $C(k,k',\theta)$  into a multipole series,
and found that only the first three even poles were significantly different from Gaussian fields.

We further decomposed these $C_{\ell}(k,k')$  matrices into diagonal terms
and cross-correlation coefficient matrices, from which we extracted the principal Eigenvectors.
This allowed us to break down  the underlying six-dimensional covariance into a set of Eigenvectors, Eigenvalues and three diagonals terms.
We provided simple fitting formulas for each of these quantities, and thus enable one to construct a full six-dimensional covariance
matrix with an accuracy at the few per cent level.

Intrinsically, non-Gaussianities introduce $N^{2}$ matrix elements to be measured in N-body simulations, as opposed to $N$
for Gaussian fields. With the proposed method, the number of parameters to measure is reduced to a handful,
even if the survey selection function is non-trivial. This opens up the possibility to measure non-Gaussianities 
directly from the data, which we will investigate in part II of our work. 

This factorization is necessary in order to estimate unbiased non-Gaussian error bars on a realistic galaxy survey, 
that must include the effect of the survey selection function.
We found that in the case of the 2dFGRS selection function, the non-Gaussian fractional variance at $k \sim 0.1 h\mbox{Mpc}^{-1}$ 
is larger by a factor of three compared to the estimate from the FKP prescription, and more than an order of magnitude at $k \sim 0.4 h\mbox{Mpc}^{-1}$.
With similar techniques, we were able to propagate a few templates of non-Gaussian Poisson error matrices into the convolution
and estimate the impact on the measured power spectrum.
We showed that with the 2dFGRS selection function, the non-Gaussian Poisson noise corresponding  to  a number density significantly lower than $0.1 h^{3} Mpc^{-3}$ has a large effect on the fractional variance at scales relevant for BAO analyses and should be incorporated in an unbiased analysis. 
The cross-correlation coefficient matrix of the convolved power spectrum  shows that the correlation propagates to larger scales in the convolution process,
and should have a larger impact on BAO analyses for instance. 
We conclude by emphasizing on the fact that constraints on cosmological parameters obtained from BAO analyses of galaxy surveys
are currently significantly biased and suboptimal, but that both of these effects can now be dealt with  in further analyses.

\section*{Acknowledgments}

The authors would like to thank Olivier Dor\'{e}, Ilana MacDonald and Patrick McDonald  for useful discussions,  Daniel Eisenstein, Martin White and
Chris Blake for their comments on early versions of the manuscript, and to
acknowledge the financial support provided by NSERC and FQRNT. The simulations and computations
were performed on the Sunnyvale cluster at CITA. 
We also acknowledge the 2dFGRS mask software written by Peder Norberg and Shaun Cole.

\appendix

\section[]{Normalization}
\label{app:norm}

\begin{figure}
  \begin{center}
   \centering
  \includegraphics[width=3.2in]{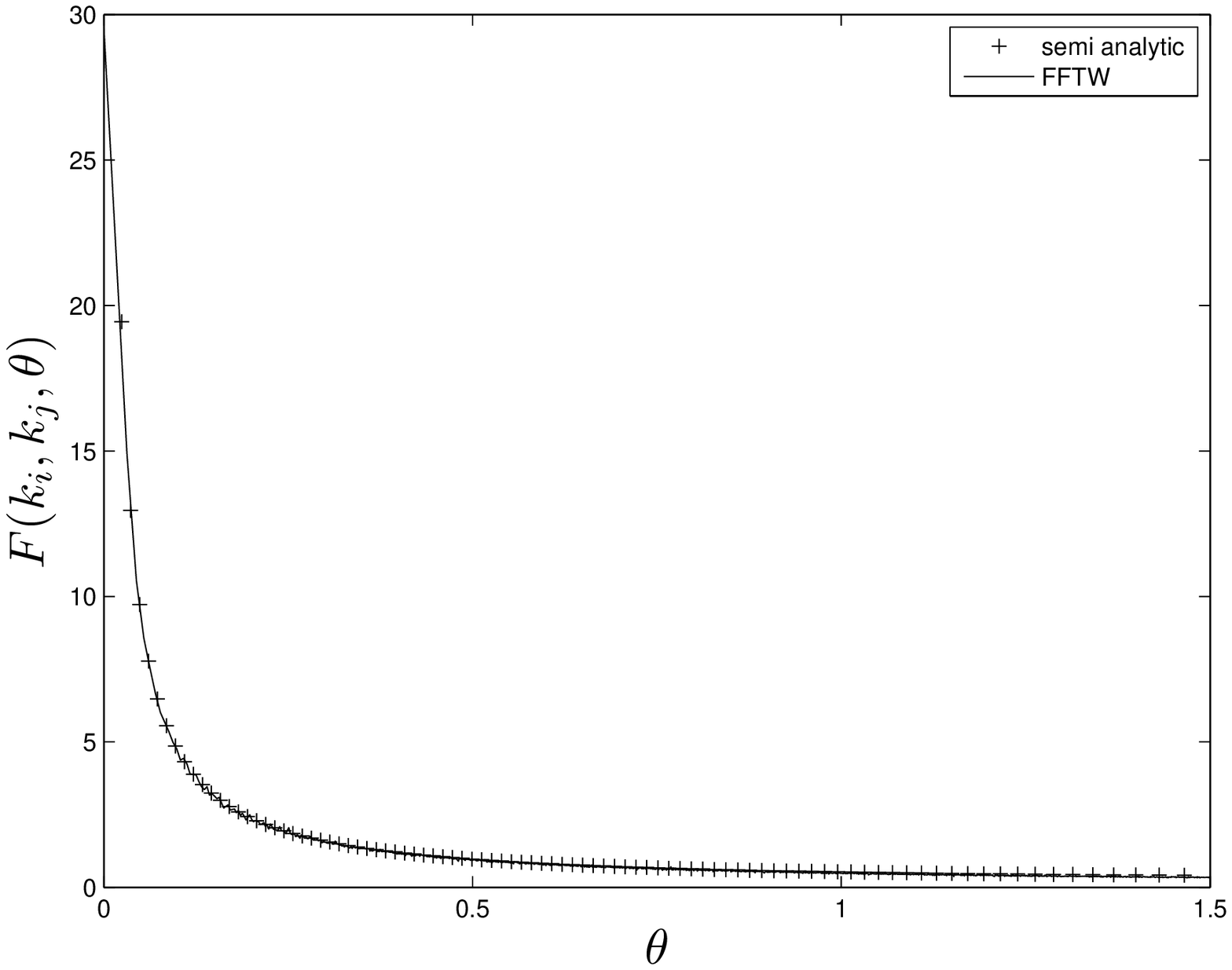}
  \caption{Function that appears in the normalization of the 
    angular covariance, $F(k_{i}, k_{j}, \Delta k)$ (equation  \ref{eq:F}),
    plotted here versus $\theta$, for two shells of $k_{i,j}=1.0h \mbox{Mpc}^{-1}$. 
    This geometrical factor represents the number of combinations, for each angle separation, as obtained 
    from a given set of shells.
    The dotted curve was obtained from the numerical integration of Eq.\ref{eq:F}, 
    and fits very well the results obtained from the {\tt FFTW} (solid curve).
    Curves like this one are obtained numerically for each scale combination $k_{i,j}$.}    
    \label{fig:Sigma_norm}
  \end{center}
\end{figure}

We present in this appendix the interpretation of the normalization $\Sigma^{ij}_{N}(\Delta k)$ 
of the angular covariance (see [Eq. \ref{eq:norm} and \ref{eq:F}]).
The shape of this function is plotted in Fig.  \ref{fig:Sigma_norm}, for a scale
that corresponds to $k = 1.00 h \mbox{Mpc}^{-1}$, which 
lies close to the transition between trans-linear and non-linear regime at $z=0.5$.

In fact, the normalization is simply a counting of the
different combinations of ${\bf k_{i,j}}$ that produce a given $\Delta{\bf k}$. 
We see from the figure that at small angles (i.e. small $\Delta k$), there is a lot of possible combinations, 
and that the function rapidly decreases with the angle. To capture this, 
imagine counting the number of times we can embed  both ends of a given vector on a given spherical shell.
In the thin shell approximation  and for a vector with non-zero length,  
the vector can be moved around an axis  parallel to the vector, 
that also passes through the centre of the shell.
In terms of solid geometry,  the vector 
spans the flat side of a cylinder that intersects the shell. 
The normalization is proportional to the circumference of the cylinder's basis,
and as the length of the vector decreases, so does the height of the cylinder,
hence its circumference increases. 
In the discrete  case, the shells have a finite thickness,
and are constructed out of a grid.
The normalization thus produces integer counts, which discretizes the  subtended angles.


\section{Legendre-Gauss Weighted Summation}
\label{app:colocation}

The conversion of the integral into a sum is performed using a Legendre-Gauss weighted sum\citep{ref:colocation}, 
in which $\ell$ `colocation' knots, which we label $\mu_{k}$ with $k=1, 2, \hdots \ell$, are placed 
at the zeros of the Legendre polynomial $P_{\ell}(\mu)$. We choose $\ell =101$, 
and we exclude the end points at $\mu = \pm 1$ in order to isolate the zero-lag contribution.
The weights $w_{k}$ are given by:
\begin{eqnarray}
  w_{k} = \frac{2}{(1-\mu_{k}^{2})(dP_{\ell=101}/d\mu(\mu_{k}))^{2}}
\end{eqnarray}
This Gaussian quadrature gives an exact representation of the integral for polynomials of degree $201$ or less,
and provides a pretty good fit to most of our $C(k_i,k_j, \theta)$.
In the linear regime, the discretization effect becomes important, and the number of angles one can make between the 
grid cells drops down as $k^{2}$. 
In the case were fewer points are available, we choose  $\ell$ = $51$, $21$, $11$ or $5$ depending on the number of available angular bins.
Once we have specified the knots, then, for each scale combination, we interpolate the angular covariance on to these knots,
and then perform the weighted sum.
As mentioned above, we always treat the zero-lag point separately in order
to avoid interpolating its value to the nearest neighbors.
We thus break the summation in two pieces:
\begin{eqnarray}
   \lefteqn{C_{\ell}^{ij} = 2\pi \sum_{\mu_{k} \ne \pm 1} P_{\ell}(\mu_{k}) C(k_i,k_j,\mu_{k}) w_{k} + 
    2\pi C(k_i,k_j, \mu=1) \Delta \mu (1+(-1)^{\ell})} \nonumber \\
\label{eq:Cl_rho_dis}
\end{eqnarray}
The factor of $2\pi$ comes from the integral over the $\phi$ angle, and 
$\Delta \mu$ is half the distance to the first knot.

\section{Eigenvector of the Poisson noise}
\label{app:PoissonEigenFit}

\begin{figure}
  \begin{center}
   \centering
  \includegraphics[width=3.2in]{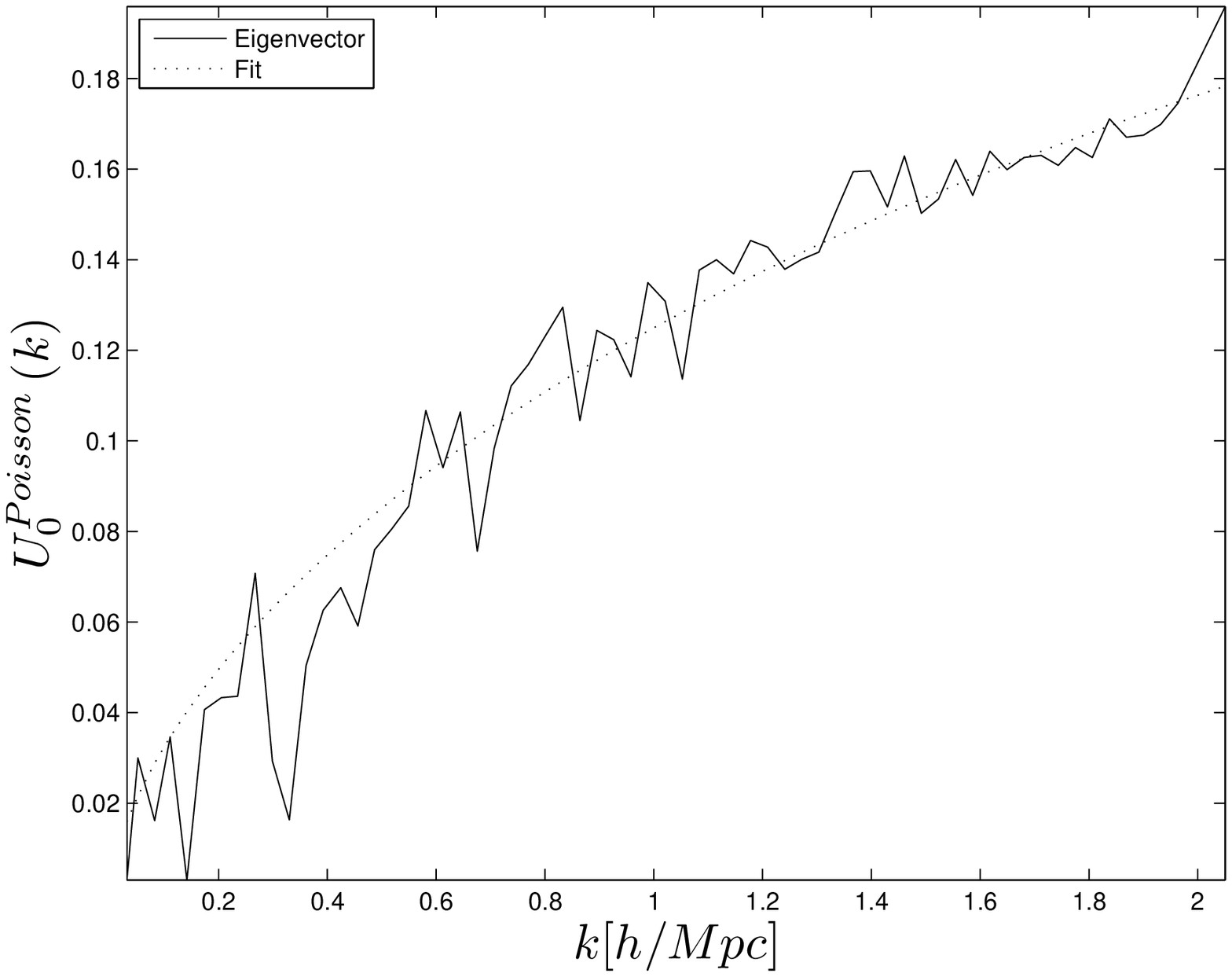}
  \caption{Principal Eigenvector of the cross-correlation coefficient matrix associated with the non-Gaussian Poisson noise,
  compare to our best-fitting formula.}    
    \label{fig:Poisson_eigenvector}
  \end{center}
\end{figure}

\begin{figure}
  \begin{center}
   \centering
  \includegraphics[width=3.2in]{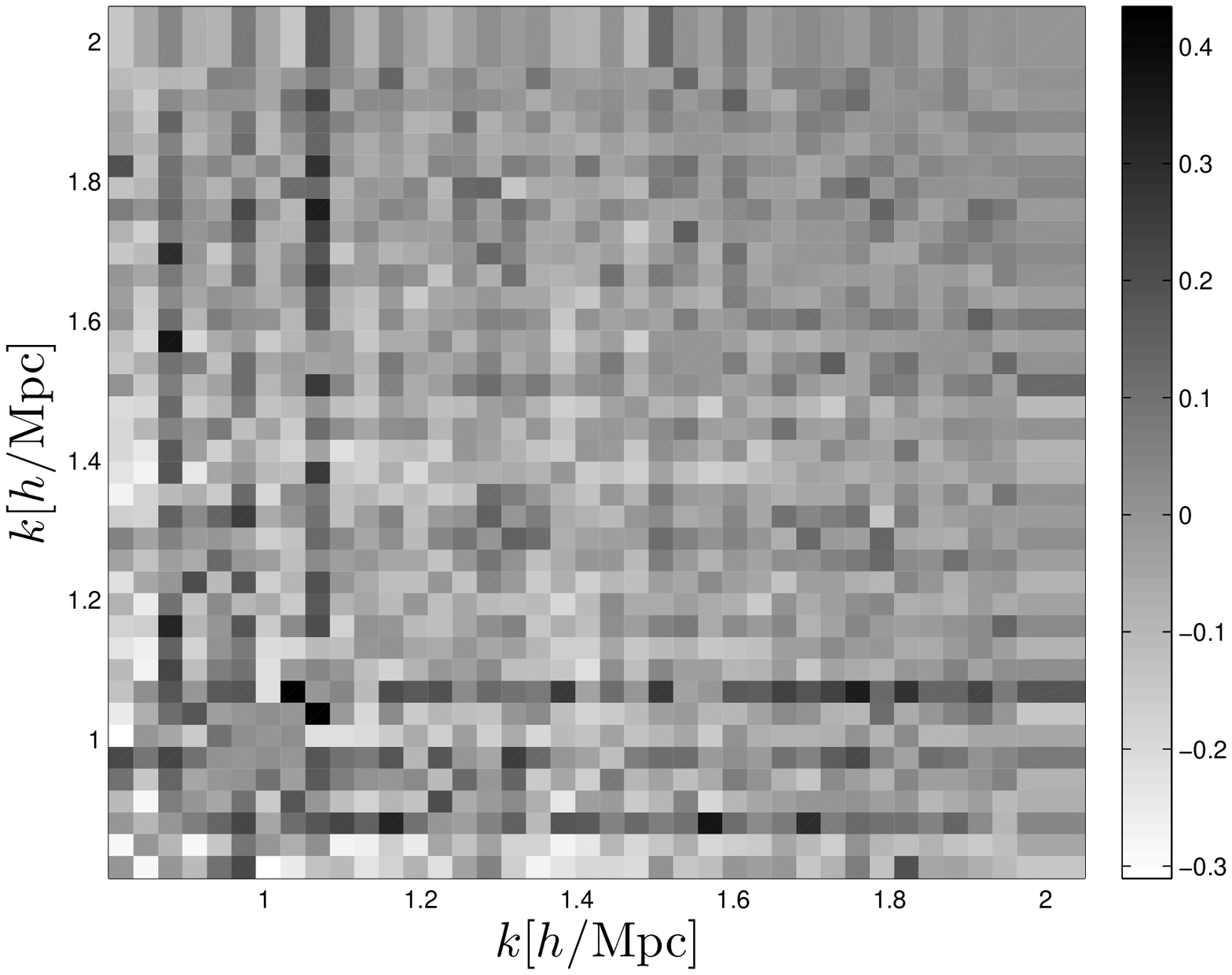}
  \caption{Fractional error between the original cross-correlation coefficient matrix associated with the non-Gaussian Poisson noise,
 and that constructed with our best-fitting functions. We have not shown the lowest $k$-modes since these are very noisy.}    
    \label{fig:frac_err_poisson_fit}
  \end{center}
\end{figure}

This Appendix presents the Eigenvector that best describes the non-Gaussian Poisson
noise, as discussed  in section \ref{subsec:non_gaussian_poisson}.
We restrict ourselves with the case where the number density is the highest,
even though similar analyses can be carried for the other values of $n$ we studied in this paper.
We present in Fig. \ref{fig:Poisson_eigenvector}  the Eigenvector itself, compared to the best-fitting formula provided.
We next  compare the covariance matrix constructed from the fitting functions with the original,
and present the fractional error in Fig. \ref{fig:frac_err_poisson_fit},
which shows  a few per cent agreement. When compared with the predictions from \citep{2006NewA...11..226C},
we observe that the overall trends are consistent: first, the Gaussian contribution to the error 
decreases as one probes smaller scales. Second, densities with lower $n$ see their Gaussian contribution 
being reduced in the trans-linear regime, where the non-Gaussian Poisson counting becomes more important. 
Third, densities with  lower $n$ produce larger cross-correlation coefficients of tran-linear scales, also
in accordance with the predictions.

%


%




\bibliographystyle{mn2e}
\bibliography{mybib3}{}

\bsp

\label{lastpage}

\end{document}